\documentclass[12pt]{article}

\usepackage{lscape}
\usepackage{multirow}
\usepackage{bm}
\usepackage{amssymb,amsmath}
\usepackage{amstext}
\usepackage{amscd}
\usepackage{graphics}
\usepackage{epsfig}
\usepackage{shadow}
\usepackage[all]{xy}
\usepackage{hyperref}

\def\ss{\scriptstyle}

\def\Real{\mathbb R}
\def\G{{\hspace{.3mm}\bf g\hspace{.3mm}}}
\def\GRAD{{\hspace{.3mm}\bf grad\hspace{.3mm}}}
\def\DIV{{\hspace{.3mm}\bf div\hspace{.3mm}}}
\def\TR{{\hspace{.3mm}\bf tr\hspace{.3mm}}}
\def\N{{\hspace{.3mm}\bf N\hspace{.3mm}}}

\def\N{{\hspace{.3mm}\bf N\hspace{.3mm}}}
\def\L{{\hspace{.3mm}\bf L\hspace{.3mm}}}
\def\La{{\hspace{.3mm}\bf \Lambda \hspace{.3mm}}}

\def\th{\theta}
\def\a{\alpha}

\def\be{\begin{equation}}
\def\ee{\end{equation}}
\def\beq{\begin{equation}}
\def\eeq{\end{equation}}
\def\bea{\begin{eqnarray}}
\def\eea{\end{eqnarray}} 

\def\eqn#1{(\ref{#1})}

\def\nn{\nonumber}
\def\coker{\operatorname{coker}}

\def\sideremark#1{\ifvmode\leavevmode\fi\vadjust{\vbox to0pt{\vss
 \hbox to 0pt{\hskip\hsize\hskip1em
 \vbox{\hsize3cm\tiny\raggedright\pretolerance10000
  \noindent #1\hfill}\hss}\vbox to8pt{\vfil}\vss}}}

\begin{document}
\thispagestyle{empty}

\vspace{.8cm}
\setcounter{footnote}{0}
\begin{center}
\vspace{-25mm}
{\Large
 {\bf BRST Detour Quantization}\\[5mm]

 {\sc \small
     D.~Cherney$^{\mathfrak C}$, E.~Latini$^{\mathfrak L}$,  and A.~Waldron$^{\mathfrak W}$\\[4mm]

            {\em\small${}^{\mathfrak C,\mathfrak W}\!$
            Department of Mathematics\\ 
            University of California,
            Davis CA 95616, USA\\
            {\tt cherney,wally@math.ucdavis.edu}\\[2mm]
          ~${}^\mathfrak{L}$ 
Dipartimento di Fisica,\\ Universit\`a di Bologna, via Irnerio 46, I-40126 Bologna, Italy 
and INFN, Laboratori Nazionali di Frascati, CP 13, I-00044 Frascati, Italy \\
  {\tt latini@lnf.infn.it}\\[2mm]}
            }
 }

\bigskip

{\sc Abstract}\\[-4mm]
\end{center}

{\small
\begin{quote}
We present the BRST cohomologies of a class of constraint (super) Lie algebras as 
detour complexes. 
By giving physical interpretations to the components of  detour complexes as gauge invariances, Bianchi identities and equations of motion we obtain a large class of new gauge theories.
The pivotal new machinery is a treatment 
of the ghost Hilbert space designed to manifest the detour structure.
Along with general results, we give details for
 three of these theories which correspond to gauge invariant spinning particle models of totally symmetric, antisymmetric and K\"ahler antisymmetric 
forms. 
In particular, we give details of our recent announcement of a~$(p,q)$-form K\"ahler electromagnetism. 
We also discuss
how our results generalize to other special geometries.
\end{quote}
}

\newpage

\tableofcontents

\newpage

\section{Introduction}

\label{intro}

The analysis of gauge theories using BRST techniques has a long history (see~\cite{Teitelboim} for a detailed account).
The aim of this Article is to generate new gauge theories
from the cohomology of the BRST operator
(in contrast to analyzing existing gauge theories using BRST techniques to  facilitate their quantization).
We  will present a BRST detour quantization ``machine'' 
which takes as its input a quantum mechanical constraint algebra along with its representation on a Hilbert space and outputs
a gauge invariant quantum field theory. Observe that our aim is {\it not} to compute the BRST cohomology but rather
to represent it as the solution space of a gauge invariant quantum field theory.

More specifically, for rank~1 constraint algebras corresponding to worldline reparameterization invariant systems, we expand 
the nilpotent BRST charge  order by order in the diffeomorphism ghost as
$
Q_{\rm BRST}=c{\cal D} + Q - M \frac{\partial}{\partial c}
$
where~$Q^2=M{\cal D}={\cal D}M$ while~$M$ and~$Q$ are built from combinations of gauge generators, ghosts and antighosts in the standard way.
In these terms, our BRST detour complex is given by
\bea\label{GeneralDetourComplex}
\cdots\stackrel{Q}\longrightarrow \coker  M \stackrel{Q}\longrightarrow \cdots\hspace{-2.6cm}
&&\hspace{-2.4cm}
\cdots\stackrel{Q}\longrightarrow {\ker} M \stackrel{Q}\longrightarrow \cdots\nn\\
&\scalebox{1.5}{$\mid$}\hspace{-.1cm}\raisebox{-0.2mm}{\underline{\hspace{2cm}\raisebox{0cm}{${\cal D}-Q M^{-1} Q$}\hspace{2cm}}}\hspace{-.15cm}\scalebox{1.6}{$\uparrow$}&
\eea
where~$M^{-1}$ is a partial inverse of the operator~$M$ which in general has a non-vanishing kernel. Giving this formula a precise
meaning requires a careful analysis of the ghost Hilbert space inspired by the Siegel's study of first quantized BRST systems~\cite{Siegel:1990zf}.
Physically this picture can be understood as follows: the kernel of the ``long'' or ``detour'' operator~${\cal D}-Q M^{-1} Q$
is the solution space of the equations of motion. The gauge invariances and any gauge for gauge invariances
are encoded by the ``incoming'' complex with differential~$Q$ and its natural quotient action on the space~$\operatorname{coker} M$. The dual, or ``outgoing'' complex where the differential~$Q$ now acts on~$\ker M$, encodes the Bianchi identities (and Bianchi for Bianchi identities) 
obeyed by the gauge invariant equations of motion. The physical analog of the above picture is therefore

\be \label{physics}
\begin{array}{c}
\cdots \stackrel{Q} {\longrightarrow}\left(\!\!\!{\tiny \begin{array}{c} \mbox{Gauge}\\[1mm]\mbox{parameters}\end{array} }\!\!\!\right)
\stackrel{Q} {\longrightarrow}
\left(\!\!\!{\tiny \begin{array}{c} \mbox{Gauge}\\[1mm]\mbox{fields}\end{array}}\!\!\!\right)
~~~~~~
\left(\!\!\!{\tiny \begin{array}{c}\mbox{Equations of }\\[1mm]\mbox{motion/currents}\end{array}}\!\!\!\right)
\stackrel{Q}{\longrightarrow} 
\left(\!\!\!{\tiny \begin{array}{c}\mbox{Bianchi/Noether}\\[1mm] \mbox{identities}\end{array}}\!\!\!\right)\stackrel{Q}{\longrightarrow} \cdots~~
\\[1mm]
\hspace{-1.3cm} \Big|\hspace{-.9mm}\raisebox{-2.2mm}{\underline{\quad \ \  \raisebox{1mm}{${\scriptstyle {\cal D}-Q M^{-1} Q}$}\quad\;}} \hspace{-1.4mm}{\Big\uparrow}
\end{array}
\ee

The notion of a ``detour complex'' wherein a detour operator connects a complex and its
dual was first introduced by conformal geometers in~\cite{Branson:2003an} where higher order, gauge invariant and Weyl covariant generalizations of Maxwell's equations were discovered by studying detours connecting the de Rham complex to its dual. An application to higher spin systems interacting with gravitational backgrounds was given in~\cite{GHW}. The detour idea was applied to higher spin systems and their worldline path integral quantization  in~\cite{Bastianelli:2009eh}. The methods presented here subsume the latter higher spin results  and should also be extendible to the conformal geometry setting.

Our work is partly motivated by the study of first quantized, spinning particle, worldline descriptions of higher spin systems.
These models provide explicit representations of the superalgebras we study here. Although many of our results do not depend on the details of underlying representations, let us mention some results in this area (the following citations
only give a guide to further reading--see also the reviews of higher spin systems~\cite{HSreviews}). The first quantized approach dates back to quantum mechanical studies  relating quantum $\frak{o}(N)$ spinning particles to spin~$N/2$ relativistic wave equations~\cite{Gershun:1979fb,Howe:1988ft}. These models were extended to describe conformal theories in~\cite{Siegel} and their path integral quantization and generalization to conformally flat backgrounds was given in~\cite{Bastianelli:2008nm} (see also \cite{Marnelius:2009uw} for a recent analysis). 

At the first quantized level, there were various further developments that are important for our current study. Firstly, one needs  models
whose quantum mechanical Hilbert space encompasses tensors with all possible symmetry types. This idea was already present in the pioneering work of~\cite{Labastida:1986gy} and mixed symmetry higher spin models have been subsequently studied in quite some detail~\cite{mixed1,Ouvry:1986dv,mixed2}.
The underlying quantum mechanical models generalizing the $\frak{o}(N)$ models to $\frak{osp}(Q|2p)$ models whose wavefunctions 
yield arbitrary tensors were given in~\cite{Hallowell:2007qk,Hallowell:2007zb} and generalized to conformal models in~\cite{Burkart:2008bq}.

The other important direction is to consider the BRST quantization of gauged, or ``spinning particle'' versions of these models.
BRST approaches to higher spins are typically characterized by a large set of auxiliary fields which correspond to what is known
as an ``unfolded approach'' to higher spins~\cite{unfold}. This correspondence is described in detail in~\cite{Barnich:2004cr}. 
In particular those works underly the idea above of expanding the BRST charge order by order in the diffeomorphism ghost which
is intimately linked with ${\cal D}$-module theory (the study of the ring of differential operators over a space). For models with purely bosonic
constraint algebras, for which the representations of the Grassmann ghost algebra are finite dimensional, that approach suffices to derive 
the corresponding long operator formul\ae~in those special cases. Of course, many of these developments hark back to String 
Field Theory~\cite{String fields}. Early  works  implementing string theoretic BRST technology for higher spin models include~\cite{Ouvry:1986dv,Bengtsson:1986ys}. Extensions to curved backgrounds were presented in~\cite{Buchbinder:2008kw}.  A short review of these BRST 
techniques can be found in~\cite{Fuster:2005eg}.
 
Modern approaches to quantization of supersymmetric theories often rely  on the pure spinor approach of Berkovits~\cite{Berkovits:2000fe,Grassi:2003cm}. There is a deep relationship between our approach and those methods. In particular, the operator $M$ in 
our BRST operator above is a certain bilinear in the ghosts, so studying its kernel can be viewed as implementing a pure spinor constraint.
A novelty of our approach, is that it converts this bilinear operator into a vector field on the ghost manifold, by way of a special 
choice of ghost polarization. It would be interesting to see if this technique could be relevant in a String Theoretic setting.

Our Article is structured as follows. In Section~\ref{BRST} we give our general BRST detour quantization theory
for a broad class of first class constraint superalgebras (specifically, maximal parabolic subalgebras of orthosymplectic 
superalgebras). These algebras look much like supersymmetry algebras,
but some of the supercharges are bosonic. In particular, they enjoy~$R$-symmetries that rotate all the ``supercharges''. In Section~\ref{R symmetry} we explain how to gauge a bosonic~$R$-symmetry generator. 
In Section~\ref{SUSY QM} we give three explicit models realizing these superalgebras, whose Hilbert spaces are symmetric tensors, differential forms and differential forms
on a K\"ahler manifold, respectively. In Section~\ref{examples} we apply BRST detour quantization to these three examples. In the first two cases we recover
the theory of massless higher spins and~$p$-form electromagnetism. For the third, we provide the details of the recently announced theory of 
K\"ahler~$(p,q)$-form electromagnetism~\cite{Cherney:2009vg}. Our Conclusions are presented in Section~\ref{conc}. There we discuss the general form of the detour
operator, plus its extension to models with other special geometries and especially quaternionic K\"ahler manifolds.

\section{BRST Detour Complexes}

\label{BRST}

Before presenting  our BRST detour quantization approach, let us briefly give a basic outline of the standard BRST approach.
Consider a first class, rank~1, quantum constraint superalgebra 
\be\label{Galg}
[G_A,G_B\}=f_{AB}^CG_C\, .
\ee
Ghosts and antighosts  (alias ghost momenta) with algebra
$$
[b_A,c^B \}=\delta_A^B\, ,
$$
are represented in a ``standard''  polarization as acting on polynomials in the ghosts with coefficients taking values in a Hilbert space
representation of the constraint algebra according to 
\be\nn
\begin{tabular}{|c|c|c|}  
\hline
~~Generator~~&~~~Ghost~~~& Anti-ghost \\
\hline
&&\\[-2mm]
$G_A$ &~$c^A$ &~$\frac{\partial}{\partial c^A}$\\[2mm]
\hline
\end{tabular}
\ee
The BRST operator
\be
\label{standardBRST}
Q_{\rm BRST}=c^AG_A-\frac 1 2 c^Cc^Af_{BC}^A\frac{\partial}{\partial c^A}~,
\ee
is then nilpotent and its cohomology equals the Lie superalgebra cohomology with coefficients in the underlying Hilbert space.
This cohomology is $\mathbb{N}$-graded by the number of ghosts and at  ghost number zero the cohomology equals~$\cap_A$ $\ker G_A$, 
{\it i.e.} the physical state space of Dirac quantization.

This approach has some obvious drawbacks. Firstly, it may not be clear how to interpret cohomologies at non-zero ghost number. 
Secondly, the {\it na\"ive} Dirac quantization of the constraint algebra may not represent the most interesting physics.
For example, in~\cite{Bastianelli:2009eh} it is shown that BRST quantization of the constraint algebra~$\{\bm \Delta, {\bf grad}, {\bf div}, {\bf tr}\}$ acting on symmetric tensors (see Section~\ref{sp2} for a detailed account of this theory) yields massless higher spin theories but at order one in the ghosts. 
This is manifestly not equal to the Dirac quantization at order zero.

Our BRST detour approach solves both these problems;  at all ghost numbers the cohomology has  physical meaning as either gauge invariances, Bianchi identities or, at zero ghost number, as equations of motion.
Indeed, for a constraint Lie {\it super}algebra
there are inequivalent definitions of
cohomology based on how these ghosts are represented~\cite{Fuchs}. For those models, one is not
simply shifting elements of cohomology from one ghost number to another, but changing the cohomology itself\footnote{We thank Albert Schwarz for this observation.}.  The choice of ghost representation given by our detour quantization is tailored precisely to the physical models we wish to study. 
Let us now give the details of our approach.

\subsection{Preliminaries}

\label{preliminaries}

Our study centers on (super) Lie algebras~$\mathfrak p$ of the form
\be
[Q_\alpha ,Q_\beta \} = J_{\alpha \beta} {\cal{D}} \,,\label{alg}
\ee
where the supermatrix~$J$ is  non-degenerate  and~${\cal{D}}$ is central.
Here the superindex~$\alpha$ takes values~$1,\dots,q+p,\ldots,q+p+2r$.
We label these algebras by their 
$R$-symmetry~$\mathfrak r =\frak{osp}(q,p|2r)$ for which the supermatrix~$J$ is an invariant tensor.\footnote{
The algebra~\eqn{alg} is the maximal parabolic subalgebra of 
~$\frak{osp}(p,q|2r+2)$~\cite{Hallowell:2007qk}. }
They can be represented by the (generalized) supersymmetry algebras of quantum mechanical models with Hamiltonian~$-2H={\cal{D}}$. (In many contexts
 this is the energy of a single particle model.) We present three explicit examples in Section~\ref{SUSY QM}. (Models for all possible algebras of this form are given in~\cite{Hallowell:2007qk}). 
 A simple but non-trivial model of this form is the~$\mathfrak{so}(1,1)$ case with algebra~$\{{\bf d},{\bf d^*} \}=\bm\Delta$. The reader may wish to keep this model in mind while reading this Section. (Indeed, typically in our models,~${\cal D}$ corresponds geometrically to the Laplacian on the manifold where the
particle moves.)

Since the (super)algebra~${\mathfrak p}$ in~\eqn{alg} is first class, we may consider its BRST quantization. Moreover, since it is a rank~1 constraint algebra, given a representation~${\cal{H}}$, BRST quantization amounts  to computing  the (super)Lie algebra cohomology~$H^{\bullet}({\mathfrak{p}},{\cal{H}} )$. Physically,~${ \cal{H} }$ corresponds to the (pre-)Hilbert space of the underlying quantum mechanical models mentioned above. 
For the calculations of this Section we do not need to specify~${\cal{H}}$. In the explicit examples given in Section~\ref{SUSY QM},~${\cal H}$ corresponds to geometric objects such as differential forms and symmetric forms on (pseudo-)Riemannian and K\"ahler manifolds.

To begin our computation, we choose a basis 
\be
\nn Q_\alpha=( Q^*_i,Q^i )\, ,
\ee 
for~${\mathfrak p}$
such that~\eqn{alg} becomes
\be
\label{alg w split}
[Q^*_i,Q^j \}=\delta_i^j {\cal{D}}\, ,
\ee
and each generator is of definite Grassmann parity. 
Here the superindex~$i$ takes on values ~$1,\dots,r+\frac{p+q+|p-q|}{2}$.
We assign the Grassmann parity 
$ \epsilon(i) \equiv {\epsilon(Q^*_i)=\epsilon(Q^i) \in \mathbb{Z}_2}$ to superindices according to the parity of their underlying supercharges (even in the case when they index ghosts whose parities are opposite to those charges). 
If~$|p-q|\neq0$, this many (super)symmetry generators obey~$Q^i=Q^*_i$. 
(Models with~$p\neq q$ are closely related to Dirac-type equations and require a separate analysis. See~\cite{AlvarezGaume:1983ig} and~\cite{Hallowell:2007zb} for discussions of these cases.)

We construct the BRST~``Hilbert space'' 
as a Fock space representation of the ghost algebra over~${\cal{H}}$ with 
the following choice of ghost polarizations:
\be\label{ghosttable}
\begin{tabular}{|c|c|c|}  
\hline
~~Generator~~&~~~Ghost~~~& Anti-Ghost \\
\hline
&&\\[-2mm]
${\cal D}$ &~$c$ &~$\frac{\partial}{\partial c}$ \\[3mm]
$Q^*_i$ &~$z^i$ &~$\frac{\partial}{\partial z^i}$\\[3mm]
$Q^i$&$\frac{\partial}{\partial p^i}$&$(-1)^{\epsilon(i)}p^i$\\[3mm]
\hline
\end{tabular}
\ee
Elements of the BRST Hilbert space are therefore formal power series in the auxiliary variables~$c,z,p$
\be
\nn {\cal{H}}_{\rm BRST}={\cal H}[[c,z,p]]={\cal H}[[z,p]] \oplus c~{\cal H}[[z,p]]\ni \Psi_{\rm BRST} = \psi(z,p)+c\chi(z,p) \, .
\ee
This space is graded by the ghost number operators
\be
\label{Ngh}
{\cal N}_{\rm gh}\equiv c\frac{\partial}{\partial c}+  z^i\frac{\partial}{\partial z^i} - p^i\frac{\partial}{\partial p^i}\equiv c\frac{\partial}{\partial c}+  N_{\rm gh}\, ,
\ee
which take integer eigenvalues {\it unbounded above and below}. This feature, implied by the ghost polarization given in (\ref{ghosttable}), is the origin of the incoming and outgoing complexes which encode gauge invariances and Bianchi identities of the models discussed below.

With these conventions the nilpotent BRST operator (which equals the  Chevalley--Eilenberg differential~\cite{ChevalleyEilenberg} in this context) reads
\be
\label{QBRST}
Q_{\rm BRST}=c\,  {\cal D} + z^i Q^*_i + Q^i \frac{\partial}{\partial p^i} - z^i\frac{\partial}{\partial p^i}\frac{\partial}{\partial c}\, .
\ee
For the ensuing computations it is useful to organize the BRST charge as an expansion in the ``diffeomorphism ghost''~$c$
\be \nn
Q_{\rm BRST}=c \, {\cal D} +Q - M\frac{\partial}{ \partial c} \, .
\ee
Here
\be
\label{MQ}
M \equiv z^i \frac{\partial}{\partial p^i }  \, ,\qquad Q \equiv  z^i Q_i^*+ Q^i \frac{\partial}{\partial p^i}\, ,
\ee
which, by virtue of the nilpotency of~$Q_{\rm BRST}$, obey the algebra 
\be
Q^2 =  M\, {\cal D} \, ,\qquad [Q,M]=0\, .\label{QM}
\ee
The operator~$M$  encodes the structure constants of the algebra~\eqn{alg} (see~\eqn{standardBRST} and~\eqn{MQ}). The ``adjoint" of~$M$ does the same, if a notion of adjoint is available.

The quantum mechanical inner product~$\langle\cdot|\cdot\rangle$ on~${\cal H}$, chosen so that the adjoint operation acts in the obvious way on~$Q^i$ and $Q^*_i$,
can be canonically extended to a bilinear function on~${\cal{H}}_{\rm BRST}$ by requiring that 
$Q_{\rm BRST}$ be ``self adjoint".
This is accomplished by defining the ``adjoint''~$*$ on ghosts as
\be
\langle z^i \phi|\theta \rangle=\langle \phi| \frac{\partial}{\partial p^{i}} \theta \rangle\, ,\qquad
\langle p^i\phi|\theta \rangle=\langle \phi| \frac{\partial}{\partial z^{i}}\theta \rangle\, ,\label{canstar}
\ee
and~$c^*=c$. This ``inner product" is degenerate;
it will be useful to consider extensions of~$\langle\cdot|\cdot\rangle$ which {\it are} genuine inner products. 
We denote by ${\mathfrak s}$ the adjoint operation associated to an arbitrary inner product 
$\langle \cdot | \cdot \rangle_{\mathfrak s}$ on~${\cal H}_{\rm BRST}$ that is equal to the standard quantum mechanical one on~${\cal H}$.
The reason that we find alternative adjoints useful is that, viewing the operator~$M$ as a linear map
\be \nn
M:{\cal H}[[z,p]]\to {\cal H}[[z,p]]\, ,
\ee
they allow us to make the following orthogonal\footnote{ 
If ~$\psi\in\ker M^{\mathfrak s}$ and~$M\alpha\in {\rm im \, M}$ 
then the two are clearly~$\langle \cdot | \cdot \rangle_{\mathfrak s}$ orthogonal.}
 decompositions of~${\cal H}[[z,p]]$ 
\bea
\label{Msdecomp}
{\cal H}[[z,p]]&=&\ker M \oplus {\rm im} M^{\mathfrak s} \, ,\\[2mm]
\label{Mdecomp}                    
{\cal H}[[z,p]]  &=&{\rm \, im} M \oplus \ker M^{\mathfrak s}\, .
\eea
Since the definition of~$M^{\frak s}$ depends on~${\frak s}$ so do such decompositions. We will denote orthogonal projections onto these spaces by subscripts. {\it E.g.}~if $\chi\in {\cal H}[[z,p]]$, then
$
\chi = \chi_{\ker M}+\chi_{{\rm im} M^{\mathfrak s}} = \chi_{{\rm im} M}+\chi_{\ker M^{\mathfrak s}} 
$,
{where~${\chi_\bullet\in \bullet}$.} Subscripts on operators denote operators composed with projections. {\it E.g.}~$Q_{\ker M} \chi$ $\equiv$  $(Q \chi ) _{\ker M}$.

Operators of the form~$Q_\bullet$ play a special role in our construction; our detour complexes are built from incoming complexes and outgoing complexes with differentials of this form. That is, much of the BRST cohomology is described by the cohomology of such operators. We now construct these complexes and the long operator connecting them.


\subsection{Detours}
\label{detours}

When acting on an arbitrary BRST state
\be \nn
\Psi_{\rm BRST}=\psi+c \chi\in{\cal H}_{\rm BRST}\, ,
\ee 
 the~$Q_{\rm{\rm BRST}}$-closure condition reads
\bea
\label{closed} 
Q\psi&=&M\chi\, ,\\[3mm]
{\cal D}\psi&=&Q \chi\, .\label{closedetour}
\eea
Since we can shift~$\Psi_{\rm BRST}$ by any exact term, these equations enjoy the invariance
\bea
\psi &\sim& \psi +Q\alpha -M\beta\, ,\label{psigauge}\\
[3mm] \chi &\sim& \chi +{\cal D} \alpha +Q\beta\, , \label{chigauge}\ 
\eea
for any~$\alpha,\beta \in {\cal H}[[z,p]]$. 

\subsubsection{Incoming Complex}\label{incoming}
Focusing on the~$\beta$ term of~\eqn{psigauge} we see that~$\psi$ is a representative of  an equivalence class~$[\psi]=[\psi+M\beta]$. 
{\it I.e.}~$[\psi]$ is an element of the cokernel of the linear map
\be \nn
M:{\cal H}[[z,p]]\to{\cal H}[[z,p]]\, .
\ee
Thanks to the identities~\eqn{QM} the operator~$Q$ is well defined and nilpotent on~$\coker M$ by setting
\be
\label{cokercleq}
Q[\phi]\equiv [Q\phi]\, ,\qquad [\phi]\in \coker M\, .
\ee
In these terms~\eqn{closed} and~\eqn{psigauge} may be stated as
\be \nn
Q[\psi]=[0] \, , \qquad  [\psi] \sim  [\psi]+Q[\alpha] \, .
\ee
This  constitutes the incoming complex
\be
\label{kerIncoming}
\shabox{
$\cdots\stackrel{Q}\longrightarrow \coker  M \stackrel{Q}\longrightarrow \cdots$}
\ee
which may be read from left to right in increasing ghost number.

Working with equivalence classes is often awkward. Therefore,
before moving on to the outgoing complex, we want to present a useful way to ``choose a gauge'' for the
transformations~$\psi\sim \psi+M\beta$. To that end we
note that the orthogonal decomposition 
${\cal H}[[z,p]]={\rm im} \, M \oplus \ker M^{\mathfrak s}$ allows one to choose the unique representative 
$\psi_{\ker M^{\mathfrak s}}$ of ~$[\psi]$ in~$\ker M^{\mathfrak s}$. 
This specifies an isomorphism between~$\coker  M$ and~$\ker M^{\mathfrak s}$
which can be viewed as  gauge fixing. In terms of this isomorphism, the closure condition and equivalence expressed in~\eqn{cokercleq} may be stated as
\be \label{incomingCohomology}
Q_{\ker M^{\mathfrak s}} \, \psi_{\ker M^{\mathfrak s}} =0 \, ,\qquad \psi_{\ker M^{\mathfrak s}} \sim \psi_{\ker M^{\mathfrak s}} +  Q_{\ker M^{\mathfrak s}}\, \alpha\, ,
\ee
where~$Q_{\ker M^{\mathfrak s}} X\equiv (Q X)_{\ker M^{\mathfrak s}}$  defines a nilpotent linear map 
${\ker M^{\mathfrak s}\to \ker M^{\mathfrak s}}$.
The  incoming complex is  therefore alternatively described as
\be 
\label{cokerIncoming}
\shabox{$
\cdots\stackrel{Q_{\ker M^{\mathfrak s}}}\longrightarrow {\rm ker} M^{\mathfrak s} \stackrel{Q_{\ker M^{\mathfrak s}}}\longrightarrow \cdots
$}
\ee
and, given a choice of adjoint~$\mathfrak{s}$, we can can produce explicit expressions for this complex.

The incoming complex makes no reference to~$\chi$, gives most of (but not all) the closure conditions on~$\psi$ and says nothing of the relationship between the two demanded by closure. 
In the next Section, we will construct an outgoing complex which reverses the role of~$\psi$ and~$\chi$ in these respects. 

\subsubsection{Outgoing Complex}
\label{Outgoing}
Whereas in the previous Section we made use of the orthogonal decomposition 
\eqn{Mdecomp}  to gauge fix~$\psi_{\ker M^{\frak s}} \in [\psi] \in {\rm coker M}$,
we now utilize the alternative orthogonal decomposition 
${\cal H}[[z,p]]=\ker M \oplus {\rm im} M^{\mathfrak s}$
to rewrite~\eqn{closed} as
\be \nn
Q\, \psi_{\ker M^{\frak s}}= M\chi_{{\rm im} \,M^{\mathfrak s}}\, .
\ee
We solve for one of these fields in terms of the other by constructing a partial inverse of~$M$; for any given adjoint~${\frak s}$ the restricted map
$$
{M: {\rm im}\, M^{\mathfrak s}\to {\rm im} \, MM^{\mathfrak s}={\rm im}\,M}
$$ 
has trivial kernel and is onto (by construction).  Thus, we may define  a partial inverse
\be  \nn
M^{-1}:{\rm im}\, M \to {\rm im}\,M^{\mathfrak s}\, .
\ee 
Moreover, when the closure conditions~$Q_{\ker M^{\frak s}}\, \psi_{\ker M^{\frak s}} = 0$ hold, we have ${Q\,  \psi_{\ker M^{\frak s}} \in {\rm  im}\, M}$  
and may therefore (unambiguously)  write
\be
\label{chiopsi} {\chi}_{{\rm im} M^{\mathfrak s}}=M^{-1}Q\ \psi_{\ker M^{\frak s}} \,. 
\ee
Having found this, we set a goal to describe the BRST cohomology without reference to this dependent field.
The {\it independent } part of~$\chi$, namely~$\chi_{\ker M}$, does not appear in~\eqn{closed};
the closure conditions on it are entirely contained in~\eqn{closedetour}. After eliminating 
${\chi}_{{\rm im} M^{\mathfrak s}}$ via~\eqn{chiopsi} that equation reads
\be
({\cal D}-QM^{-1}Q) \ \psi_{\ker M^{\frak s}}=Q{\chi}_{\ker M} \label{detclo} .
\ee

The right and left hand sides of this equation are independently zero 
if one makes a particular choice\footnote{This choice amounts to a choice of
$\coker  M$ representatives.} of the adjoint~$\mathfrak{s}$. Namely, we choose the involution corresponding to the standard Fourier transform in the ghost and antighost variables
\be
\label{sadj}
z^{i \mathfrak s}=\frac{\partial}{\partial z^{i}} \, ,\qquad
 p^{i\mathfrak s}= \frac{\partial}{\partial p^{i}}\, .
\ee
With this choice 
\be \nn
M^{\mathfrak s}=p^i\frac{\partial}{\partial z^i}\, 
\ee
and explicitly, 
\bea \nn
\ker M~&=&{\cal{H}}[[z,y]]  \\[2mm]\nn
\ker M^{\mathfrak s}&=&{\cal{H}}[[y,p]] \, ,
\eea
where~$y$ denotes the composite variables
\be
\label{ys}
y^{ij}=(-1)^{\epsilon(i)} ( p^{i} z^{j}- z^{i}p^{j} )   \, .
\ee
These ghost bilinears obey
\bea
y^{ij}\ \ &=&\ \ y^{(ij]}\, ,\nn\\[2mm]
[M, y^{ij}]\ =&0& =[M^{\frak s}, y^{ij}]\ \, ,\nn\\[2mm]
[M,M^{\frak s}] ~ &=&\ \ N_{\rm gh}\, , \label{MMN}\\[2mm]
[N_{\rm gh} , y^{ij}]&=&\ \ 0 \ ,  \nn \,  
\eea
where~$(\cdots]$ denotes unit weight symmetrization for Bose superindices and antisymmetrization for Fermi ones; $y^{ij}=(-1)^{\epsilon(i)\epsilon(j)}y^{ji}$.
This utilitarian choice of ghost number reversing adjoint is highly fortuitous: in later Sections we will show that while $M$ and $M^{\frak s}$ encode the structure constants of the algebra~\eqn{alg} (compare~\eqn{MQ} and~\eqn{standardBRST} ) the $y^{ij}$ specify mutually (super)commuting $R$-symmetries.  The latter fact has motivated the choice of overall sign (and resultant index symmetry) in~\eqn{ys}. 

We see that~$\ker M$ and~$\ker M^{\mathfrak s}$  are subspaces of 
${\cal{H}}[[z,p]]$ with ghost number~$\geq 0$ and~$\leq 0$, respectively. Since~$Q$ raises ghost number by one and~$({\cal D}-QM^{-1}Q)$ is ghost number neutral,
the left hand side of~\eqn{detclo} is an object with ghost number~$\leq 0$ whereas the right hand side has ghost number~$>0$. We conclude
\bea
Q_{\ker M^{\frak s}}\,  \psi_{\ker M^{\frak s}} \ \ &=&0\, ,\nn \\[2mm]
({\cal D}-QM^{-1}Q)\,  \psi_{\ker M^{\frak s}}&=&0\, , \label{longeq}\\[2mm]
Q \chi_{\ker M}\ \ \ &=&0  \nn \, .
\eea
The first of these equations is the closure condition for the incoming complex of Section~\ref{incoming},
the second is the detour closure condition that we discuss in detail in Section~\ref{detour} while the third is the closure condition
we now utilize to construct the outgoing complex.

Looking back at~\eqn{psigauge} 
we recall that we used~$M\beta$ to fix~$\psi_{\ker M^{\frak s}}$ which still  leaves~$\beta_{\ker M}$ free to appear in~\eqn{chigauge} as
\be \label{D?}
\chi_{\ker M} \sim \chi_{\ker M} +{\cal D} \alpha+ Q\beta_{\ker M}  \, .
\ee
To preserve any gauge choice one might make for the cohomology of the incoming complex~\eqn{incomingCohomology}, we demand~$Q_{\ker M^{\frak s}}\alpha=0$ in the above equation. For this reason we can rewrite (using~\eqn{QM} and~\eqn{Msdecomp})
$$
{\cal D}\alpha=M^{-1}Q^2 \alpha=M^{-1}Q[(Q\alpha)_{{\rm im} M}+(Q\alpha)_{\ker M^{\frak s}}]=M^{-1}QM\phi=Q\phi
$$
for some~$\phi$ such that~$M\phi=(Q\alpha)_{{\rm im }M}$. 
Thus,~\eqn{D?} amounts to
\be \label{noD}
\chi_{\ker M} \sim \chi_{\ker M} +(Q\beta)_{\ker M}  \, .
\ee
Moreover, thanks to the identities~\eqn{QM} 
the map
\be  \nn
Q:\ker M\to\ker M
\ee
is well defined and nilpotent. This defines the outgoing complex
\be
\shabox{
$\cdots\stackrel{Q}\longrightarrow {\rm ker} M \stackrel{Q}\longrightarrow \cdots$
}
\label{outgoing}
\ee
which imposes the closure condition given in the third line of~\eqn{longeq} with the invariance given in~\eqn{noD}. 

We have constructed two complexes,~\eqn{incoming} and~\eqn{outgoing}, both with boundary maps~$Q$, and both read from left to right in increasing ghost number. The first, the incoming complex, imposes a condition on~$\psi$. The second, the outgoing complex, imposes a condition on~$\chi$.
In the next Section we construct a long operator connecting these complexes at ghost number zero by imposing the part of the closure conditions~\eqn{closed},~\eqn{closedetour} which relate~$\psi$ and~$\chi$.

\subsubsection{The Detour}\label{detour}
 
We exhausted the information contained in~\eqn{closed} by constructing the incoming complex and by  finding~$\chi_{{ \rm im} \, M^{\frak s}}=M^{-1}Q\psi_{\ker M^{\frak s}}$.
So, it is natural to ask what information {\eqn{closedetour} contains beyond that already given in~\eqn{closed}.
Multiplying~\eqn{closed} by~$Q$ on both sides yields
\be  \nn
M({\cal D} \psi_{\ker M^{\frak s}}-Q\chi)=0\, ,
\ee
which is almost equivalent to~\eqn{closedetour} in the following sense: applying the partial inverse 
$M^{-1}:{\rm im} M \to {\rm im} M^{\frak s}$ 
we learn
\be  \nn
\Big[{\cal D} \psi_{\ker M^{\frak s}}-Q\chi\Big]_{{\rm im} \, M^{\mathfrak s}}=0\, .
\ee
Because of the orthogonal decomposition~\eqn{Msdecomp} we may say that the {\it new } information in~\eqn{closedetour} is 
\be \nn
\Big[{\cal D} \psi_{\ker M^{\frak s}}-Q\chi\Big]_{\ker M}
=0.
\ee
In the previous Section we showed that the outgoing closure condition was $Q\chi_{\ker M}=0$ and that~$\chi_{{\rm im}M^\frak{s}}=M^{-1}Q\psi_{\ker M^{\frak{s}}}$. With these substitutions we discover that the remaining closure condition is
\be \nn
[({\cal D}- Q M^{-1} Q ) \psi_{\ker M^{\frak s}}]_{\ker M}\, =0.
\ee
The projection onto ${\ker M}$ acts as the identity thanks to~\eqn{QM}.

Therefore,~${ G}={\cal D}- Q M^{-1} Q~$ is the long operator. 
Indeed, because~${ G}M=M { G}=0$ we may say that~${ G}$ is a map~${\coker M \to \ker M}$.
Further,~\eqn{QM} guarantees that~$GQ_{\ker M^{\frak s}}=QG=0$ allowing us to construct the detour complex 

\bea
\cdots\stackrel{Q_{\ker M^{\frak s}}}{-\!\!\!-\!\!\!\longrightarrow} \stackrel{}{~\ker M^{\frak s}~}
\stackrel{Q_{\ker M^{\frak s}}}{-\!\!\!-\!\!\!\longrightarrow} \stackrel{}{~\ker M^{\frak s} } 
{ \to 0} \hspace{-2cm}
&\qquad&\hspace{-2cm}
0\to\stackrel{}{~{\ker} M ~}
\stackrel{Q}{-\!\!\!-\!\!\!\longrightarrow} \stackrel{}{~{\ker} M~ }
\stackrel{Q}{-\!\!\!-\!\!\! \longrightarrow} \cdots
\nn\\
&\scalebox{1.5}{$\mid$}\hspace{-.22cm}\raisebox{-.1cm}
{ \underline{ \hspace{1.5cm}\raisebox{.1cm}{\bf ~G}\hspace{1.5cm}} }
\hspace{-.29cm}
\scalebox{1.8}
{$\uparrow$}&
\eea

The detour complex above is has an additional desirable feature: the incoming and outgoing complexes are naturally read from left to right in increasing ghost number. 
The long operator 
preserves ghost number $N_{\rm gh}$ 
and may be restricted to ghost number~$0$ without loss of information. 
It is, then, a map from~${\rm ker} M^{\frak s}$, a space whose elements have ghost number~$\leq0$, to ~${\rm ker} M$, a space whose elements have ghost number~$\geq 0$. So,~${ G}$ must have trivial action away from ghost number 0. Without loss of closure information we redefine
\be \nn
{ G} :  \ker M^{\frak s} |_{N_{\rm gh}=0}\to \ker  M|_{N_{\rm gh}=0}\, .
\ee
Actually, because of ~\eqn{MMN} we have the identity\footnote{ 
$\ker M^s \cap \ker M \!\supset\! \ker M|_{N_{\rm gh}=0}$ 
since, if
$M\psi=0$
and
$N\psi=MM^*\psi-M^*M\psi=0$ then
$MM^* \psi=0$ and so,
applying our partial inverse,~$M^*\psi=0$.
} 
\be \nn
\ker M^{\frak s} \cap \ker  M =\ker M^{\frak s} |_{N_{\rm gh}=0}= \ker  M|_{N_{\rm gh}=0}
\ee
and therefore
\bea \nn
{ G} :  {\ker M^{\frak s}} \cap \ker  M \to  \ker M^{\frak s} \cap \ker  M \, .
\eea
We see that the long operator acts on the intersection of the incoming and outgoing complexes at ghost number zero. This clears up the issue raised above~\eqn{chiopsi}: $M^{-1}$ is now always well defined in the long operator. This is because $Q_{\ker M^{\frak s}}$ acts trivially at ghost number zero in the incoming complex (because there are no ghost number $1$ fields in $\ker M^{\frak s}$) and thus, $Q$ maps this space to the image of $M$, where $M^{-1}$ is defined.

The detour complex is written with more detail as
\bea 
\cdots\stackrel{Q_{\ker M^{\frak s}}}{-\!\!\!-\!\!\!-\!\!\!\longrightarrow} \stackrel{-1}{\ \ker M^{\frak s}\ }
\stackrel{Q_{\ker M^{\frak s}}}{-\!\!\!-\!\!\!\longrightarrow} \stackrel{0}{\ \ker M^{\frak s}\  } 
\stackrel{\phantom{Q}}{ \to 0\ \ \ \ } \hspace{-2.6cm}~
&&\hspace{-2.3cm}
\ \ \ 0\to \, \stackrel{1}{\ {\ker} M \ }
\stackrel{Q}{-\!\!\!-\!\!\!\longrightarrow} \stackrel{2}{\ {\ker} M \ }
\stackrel{Q}{-\!\!\!-\!\!\!\longrightarrow} \cdots
\nn\\ 
&\scalebox{1.4}{$\mid$}\hspace{-.2cm}\raisebox{-.1cm}
{ \underline{ \hspace{1.2cm}\raisebox{.1cm}{\bf ~G}\hspace{1.5cm}} }
\hspace{-.3cm}
\scalebox{1.6}
{$\uparrow$}&
\label{detcomplexfull}
\eea
Here, the integers above the modules indicate eigen-subspaces of~${\cal N}_{\rm gh}$ as defined in~\eqn{Ngh}.

This concludes our general construction of the BRST detour complex associated with the algebra~\eqn{alg}. Our choices of ghost polarization\eqn{ghosttable} and of decompositions~\eqn{Msdecomp},~\eqn{Mdecomp}  allowed us display the BRST cohomology in a revealing way, detailing gauge invariances and Bianchi identities of an equation of motion as in~\eqn{physics}.  
An interesting feature of the detour cohomology at ghost number zero is that it still depends on ghosts
through the  bilinears~$y^{ij}$. (In a ``straight polarization'' approach, the ghost number
zero cohomology is necessarily ghost-free.)
However, as we show in the following Section, the~$y^{ij}$ correspond to~$R$-symmetry transformations of the algebra~\eqn{alg}. 
There we present a general method for gauging a single~$R$-symmetry in addition to the generators of~\eqn{alg} which removes the lingering  ghosts. We discuss the procedure for gauging multiple $R$-symmetries in~\cite{CLW}.

\subsubsection{Gauging an $R$-Symmetry}
\label{R symmetry}
If we add to the algebra~\eqn{alg} a single bosonic~$R$-symmetry\,generator\,$\tau$\,acting\,as
\bea \nn
[\tau,Q_\alpha ]=\tau^{~\beta}_\alpha Q_\beta \, ,
\eea
we again obtain a closed, first class, rank one algebra. 
We introduce the odd ghost~$e$ corresponding to~$\tau$ and the extended BRST Hilbert space 
$${\cal H}[[c,z,p,e]]={\cal H}[[c,z,p]]\oplus e{\cal H}[[c,z,p]] \, .$$ 
The Chevalley--Eilenberg differential of this new algebra is
\bea \nn
Q_T=Q_{\rm BRST}+ T \frac{\partial}{\partial e}\, ,
\eea
where   the operator 
$T=\tau- t$ with~$t$ a bilinear in~$\{z,p,\frac{\partial}{\partial z}, \frac{\partial}{\partial p} \}\equiv \{c^\alpha\}$ is defined by
\bea \nn
[t,c^\alpha ]=\tau^{{~}\alpha}_\beta c^\beta \, , ~~[M,t]=0  \, .
\eea
Making use of the definition~\eqn{MQ} these conditions demand  
\be \label{ghostsym}
[Q,\tau ]= [Q,t ] \Rightarrow [T,Q]=0\, ,
\ee
and that the structure constants, encoded by~$M$, are preserved by~$t$.
This guarantees that~$Q_T$ is 
nilpotent.

Since\footnote{See also Theorem 8.3 of~\cite{Teitelboim}.} $\{Q_{\rm BRST},T\frac{\partial}{\partial e}\}=0$, the cohomology of $Q_T$ is equal to the cohomology of $Q_{\rm BRST}$ over $H^{\bullet}(Q_{T})={\coker T}   \oplus  e  \, {\ker} \, T $.  
That is to say, the cohomology of~$Q_{T}$ is the sum of the cohomology of
~$Q_{\rm BRST}$ acting on~${\coker} \, T$ and the cohomology 
of~$Q_{\rm BRST}$ acting on~$\ker T$. 
We label these two sectors by their~$R$-symmetry ghost numbers 0 and 1, respectively. 
The rest of the cohomology problem is solved by  following the procedure of the previous Section restricted to these subspaces.  That is, we obtain two detour complexes of the form presented in~\eqn{detcomplexfull}. Explicitly, they are

\bea
\label{kerRBcomplex}
\cdots
\stackrel{Q}\longrightarrow \coker \,  M \cap {\rm coker} \,  T
\stackrel{Q}\longrightarrow \cdots \hspace{-2.6cm}
&&\hspace{-2.4cm}
\cdots \stackrel{Q}\longrightarrow {\ker} M \cap {\rm coker} T
\stackrel{Q}\longrightarrow \cdots\nn\\
&\scalebox{1.5}{$\mid$}\hspace{-.1cm}\raisebox{-.1cm}{\underline{\hspace{2cm}\raisebox{.1cm}{ G}\hspace{2cm}}}\hspace{-.15cm}\scalebox{1.5}{$\uparrow$}&\\[10mm]
\label{kerR*Bcomplex}
\cdots
\stackrel{Q}\longrightarrow \coker \, M \cap \ker T
\stackrel{Q}\longrightarrow \cdots \hspace{-2.6cm}
&&\hspace{-2.4cm}
\cdots \stackrel{Q}\longrightarrow {\ker} M \cap \ker T
\stackrel{Q}\longrightarrow \cdots\nn\\
&\scalebox{1.5}{$\mid$}\hspace{-.1cm}\raisebox{-.1cm}{\underline{\hspace{2cm}\raisebox{.1cm}{ G}\hspace{2cm}}}\hspace{-.15cm}\scalebox{1.5}{$\uparrow$}&
\eea

The orthogonal decomposition 
${\cal H}= {\rm im}\,T \oplus {\rm ker}\,T^*$ implies that every 
${\rm coker} \,T$ class has a representative in 
${\rm ker}\, T^*$. 
However, since the extension of the inner product~\eqn{canstar} is degenerate, there are many such representatives differing by an element of $\ker T^*\cap{\rm im} \,T$. We deal with this ambiguity in a case by case basis in the models below.
%




Corresponding to each ghost bilinear~$y^{ij}$ defined in~\eqn{ys} there is an ~$R$-symmetry~$\rho^{ij}$ satisfying the relationship~\eqn{ghostsym}
\bea\nn
[y^{ij},Q ]=[\rho^{ij}, Q ] \, .
\eea 
Explicitly, these rotate the (super) charges as
\bea\nn
 [\rho^{ij},Q^k ]=0~, ~[\rho^{ij},Q^*_k ]=-\delta_k^i Q^j-(-1)^{\epsilon(i)} \delta^j_k Q^i \, .
\eea
It is  particularly useful to gauge a bosonic $R$-symmetry of this kind; it was shown in Section~\ref{Outgoing} that the incoming and outgoing complexes are subspaces of 
\bea\nn
\ker M~&=&{\cal{H}}[[z,y]]  \\[2mm]
\nn
\ker M^{\mathfrak s}&=&{\cal{H}}[[y,p]] 
\eea
and that the long operator acts on the intersection of these, ${\cal{H}}[[y]]$.
In the sector~$0$ detour complex the ${\rm coker} (\rho^{ij} -y^{ij})$ relations
\bea\nn
[\psi]=[\psi+\rho^{ij} \alpha_{ij} - y^{ij} \alpha_{ij}]
\eea 
may be utilized to pick $y^{ij}$ free representatives of each ${\rm coker} $ class, a process we refer to as ``ghostbusting''. Similarly, $y^{ij}$ free representatives may be chosen at each level of the sector 0 complex. In our examples, sector 1 will always be empty. For cases where multiple $R$-symmetries are gauged, we refer to a forthcoming publication~\cite{CLW}.

In the models presented below we find that the~$R$-symmetry generator~$\rho_{ij}$ corresponding to~$y^{ij}$ has a nice geometric interpretation which we hope to generalize to other models. 
The following Section provides a brief review of three models with generalized (super)symmetry algebras of the form~\eqn{alg} focusing on their geometric interpretations.


\section{Supersymmetric Quantum Mechanics} \label{SUSY QM}

As explained  in~Section~\ref{preliminaries} we refer to an algebra of the form~\eqn{alg} by its maximal $R$-symmetry algebra. 
In this sense we now present models, corresponding to~$\mathfrak{sp}(2)$,~$\mathfrak{so}(1,1)$ and~$\mathfrak{so}(2,2)$, which will yield (higher spin) field theories of totally symmetric, antisymmetric and K\"ahler  antisymmetric forms, respectively. The adroit reader, cognizant of the three quantum mechanical models reviewed here can skip directly to their BRST detour quantizations given in Section~\ref{examples}.
A detailed description of the~$\frak{sp}(2)$ and~$\frak{so}(1,1)$ theories on Riemannian manifolds can be found in~\cite{Bastianelli:2009eh},
while an elegant account of the K\"ahler theory is given in~\cite{FigueroaO'Farrill:1997ks} (for mathematical background on K\"ahler geometry, see~\cite{Griffiths}).


\subsection{$\mathfrak{sp}(2)$: Symmetric Tensors}
\label{sp2}
In flat backgrounds, models of parallel transport of a complex tangent vector along geodesics exhibit a high degree of symmetry. Remarkably, the same degree of symmetry is also found in such models on locally symmetric spaces. 
This is achieved  by curvature couplings between the
parallel transport and geodesic equations~\cite{Hallowell:2007zb}.

The symmetries of these single particle quantum mechanical models, upon quantization,
underly the algebra of differential geometry operators acting on symmetric tensors discovered long ago by Lichnerowicz~\cite{Lichnerowicz:1961}. 
We take this system as our first example of detour BRST quantization.

Let~$(M,g_{\mu\nu} )$ be a~$d$-dimensional flat (pseudo)-Riemannian manifold with coordinates~$x^\mu$. We introduce a pair of complex (Grassmann even) variables~$(z^*_\mu,z^\mu)$ and a first order   action principle
\be \nn
S^{(1)}=\int dt\Big[ \dot{x}^\mu p_\mu+iz^*_\mu\dot{z}^\mu-\frac 1 2\,  p^\mu p_\mu\Big]\, ,
\ee
whose equations of motion imply that~$x^\mu(t)$ is a (parameterized) geodesic along which the covectors~$z_\mu(t)$ and~$z^*_\mu(t)$ are parallely transported.

It is possible to gauge any combination of 
symmetries whose generators form a first class constraint algebra. We will always, at the very least, gauge worldline translations and the bosonic analog of the supersymmetry-like charges. In that case, after a Legendre transformation,  we obtain a 
second order gauge invariant action with 
 gauge multiplet~$(e,\upsilon^*,\upsilon)$: 
$$
S=\int dt\Big[\frac{1}{ 2e} (\dot{x}^\mu-i[\upsilon^*z^\mu+\upsilon z^{*\mu}]) g_{\mu\nu}(\dot{x}^\nu-i[v^*z^\nu+vz^{*\nu}])+iz^*_\mu\dot{z}^\mu\Big]\, .
$$
Analogous second order, gauge invariant, classical worldline  actions can be constructed for all the models whose BRST detour quantizations
are studied in this paper.
The BRST detour quantization of this extremely simple gauge invariant model will produce a wide range of (linearized) physical quantum 
field theories. In essence, this point was first observed by Labastida quite some time ago in~\cite{Labastida:1986gy}. Those results follow almost immediately
from our BRST detour machinery as we shall show in Section~\ref{sp2BRST}.

The symplectic structure, and in turn quantum commutators of the model, may be read directly from the  first order action 
$$
[x^\mu,p_\nu]=i\delta^\mu_\nu~,~~~~~~~~[z^\mu,z^*_\nu]=\delta^\mu_\nu \, .
$$ 
Guided by geometrical intuition,  we represent this algebra on the Hilbert space 
${\cal H}=L^{2}({M} ) \otimes \mathbb{C}[[z^{*}]]$ 
({\it i.e.} the tensor product of the~$L^{2}$ representation of the~$(x,p)$, 
and  Fock representation of the~$(z^{*},z)$, Heisenberg algebras) and make the operator identifications
\be\begin{array}{ccc}
&p_\mu\to -i\frac{\partial}{\partial x^\mu}&\nonumber\\[3mm]
z^{\mu}\to dx^\mu &&  z^*_\mu\to \frac{\partial}{\partial(d x^\mu)} \, .
\end{array}
\nn
\ee
Wavefunctions are then sections of the symmetric tensor bundle over~$M$
\be  \nn
\Psi(x,dx)=\sum_{s=0}^{\infty}\psi_{\mu_{1}...\mu_s}dx^{\mu_{1}}...\,dx^{\mu_s}\in \Gamma(\odot T^*M)\equiv {\cal H}\, .
\ee 
The bosonic quantum ``supercharges", when translated into this geometrical language, are
$$
{\bf grad}=dx^\mu\partial_\mu\, ,~~~~~~~~~~~~~~~~~~~~~~{\bf div}=\frac{\partial}{\partial (dx_\mu)}\partial_\mu\, ,
$$
and correspond to the symmetrized gradient and divergence, respectively. T he quantum Hamiltonian reproduces the Bochner Laplacian operator
$$H\equiv -\frac 1 2\ {\bf \square}~.$$ 
Together, these form the algebra 
\be  \nn
[{\bf div},{\bf grad} ]={\bf \square}\, ,
\ee
so identifying~$Q_\alpha=({\bf grad}, {\bf div})$ and~${\cal D}=\square$, we obtain the algebra~$\mathfrak p$ in~\eqn{alg} with~$p=q=0$ and~$r=1$.

The generators of the~$R$-symmetry group~$\mathfrak{sp}(2)$ act on~$\Psi$ as the symmetric form degree, metric multiplication and trace operators: 
\be  
\begin{array}{rclrclrcl}
&&&{\bf N}&=&dx^\mu\frac{\partial}{\partial (dx^\mu)}\, ,&&&\nonumber\\[3mm]
{\bf g}&=&dx^\mu g_{\mu\nu}dx^\nu\, ,&&&&{\bf tr}&=&\frac{\partial}{\partial (dx^\mu)} g^{\mu\nu}\frac{\partial}{\partial (dx^\nu)}~,
\end{array}
\ee
$$
{\bf [N,tr]}=-2\ {\bf tr}\, ,\qquad [{\bf tr},{\bf g}]= 4{\bf N}+2 d\,  ,\qquad [{\bf N},{\bf g} ]=2\ {\bf g}\, .
$$
This algebra acts upon the ``supercharges"~$({\bf grad}, {\bf div})$ as an~$R$-symmetry doublet:
$$
[{\bf tr},{\bf grad} ]=2\ {\bf div}\, ,\quad
\Big[{\bf N},\left(\!\!\begin{array}{c}{\bf grad}\\{\bf div}\end{array}\!\!\right)\Big]=\left(\!\!\begin{array}{c}{\bf grad}\\-{\bf div}\end{array}\!\!\right) \, , \quad
[{\bf div},{\bf g} ]=2\ {\bf grad} \, .
$$
The algebra generated by~$\square,{\bf grad, div}$ and~${\G}$
is a rank~1 constraint algebra whose  cohomology will be discussed in Section~\ref{sp2BRST}. 


\subsection{$\mathfrak{so}(1,1)$: Differential Forms}
\label{so11}
We again let~$({M},g_{\mu\nu})$ be a Riemannian manifold of dimension~$d$, but now take~$g_{\mu\nu}$ to be an arbitrary metric. The 
complex tangent vectors~$(z^\mu,z^{*\mu})$ of the~$\mathfrak{sp}(2)$ model are replaced by a pair of Grassmann, vector-valued variables 
$(\th^\mu,\th^{*\mu})$ which will form a doublet under an ~$\mathfrak{so}(1,1)$~$R$-symmetry. The model we are about to describe will have differential
forms as wavefunctions and is precisely the~${\cal N}=2$ supersymmetric quantum mechanical model employed by Witten to describe Morse theory~\cite{Witten:1982im}; the path integral quantization of this system have been discussed in~\cite{Bastianelli:2005uy}.

The  first order  action principle of the model reads:
\be
\label{so11h}
S^{(1)}=\int dt\Big[\dot{x}^\mu \pi_\mu+i \th_\mu^*\ \frac{ \nabla \th^\mu}{d t}-\frac 1 2 \pi_\mu g^{\mu\nu}\pi_\nu+\frac 1 2 \  R^\mu{}_\nu{}^\rho{}_\sigma~\th^*_\mu\th^\nu\th^*_\rho\th^\sigma\Big]\, .
\ee
 This action is invariant under rigid worldline translations,~${\cal N}=2$ supersymmetry and an~$\mathfrak{so}(1,1)$~$R$-symmetry under which the supercharges form a doublet representation.

The symplectic structure and quantum (anti)commutation relations  follow directly form the first order action by noting that~$\pi_\mu = p_\mu + i\Gamma_{\mu\nu}^\rho\theta_\rho^*\theta^\nu$ is the covariant momentum so that
$$
[p_\mu,x^\nu]=-i\delta_\mu^\nu~,~~~~~~~~~\{\th^\mu,\th^*_\nu\}=\delta^\mu_\nu~.
$$ 
Again, guided by geometrical intuition, we represent this algebra on the Hilbert 
space 
$L^2 ( { M}) \otimes {\mathbb C}[[\th^* ]]$ ({\it i.e.} the tensor product of the 
$L^2$ 
representation of the 
coordinate algebra~$[x^\mu , p_\nu] = i\delta^\mu_\nu$ and the Fock representation of the~$(\th^*,\th)$ algebra) and make the operator identifications 
\bea
&i\pi_\mu\to\nabla_\mu\, ,&  \nonumber\\[2mm]
\th^{*\mu}\to dx^\mu\, ,&&\th_\mu\to\frac{\partial}{\partial(dx^\mu)} \, .  \nn
\eea
Here, since~$(\theta^*,\theta)$ are Grassmann variables, the coordinate differentials~$dx^\mu$ {\it anticommute}
and~$\partial/\partial(dx^\mu)$ is a Grassmann derivative. In turn wavefunctions are differential forms
\be  \nn
\Psi=\Psi(x,dx)=\sum_{k=0}^{d}\psi(x)_{\mu_1...\mu_k}dx^{\mu_1}\wedge...\wedge
 dx^{\mu_k}\in \Gamma\Lambda M = {\cal H}\, .
\ee 
At the quantum level, the supercharges 
$Q=i\th^\mu\pi_\mu$ and~$Q^*=i\th^{*\mu}\pi_\mu$ act on~$\Psi$ as the exterior derivative~${\bf d}$ and its dual~${\bf d^*}$, respectively, while the quantum Hamiltonian reproduces the (curved space) form Laplacian
~$$
 -2 H\Psi=\bm\Delta \psi~.
~$$
The internal~$R$-symmetry generator~${\bf N}$, with a suitable quantum ordering,  counts form degree:
~$$
{\bf N}\ \psi_{\mu_1...\mu_k}dx^{\mu_1}\wedge...\wedge dx^{\mu_k}
 =k\ \psi_{\mu_1...\mu_k} dx^{\mu_1}\wedge...\wedge dx^{\mu_k}~.
~$$ 
The~$\mathfrak{so}(1,1)$-extended supersymmetry algebra is easily computed 
 \bea
&{\bf dd^*}+{\bf d^* \!d}=\bm\Delta\, ,&\nonumber\\[4mm]
[{\bf N},{\bf d}]={\bf d}\, ,~~~~~&&~~~~~[{\bf N},{\bf d^*}]=-{\bf d^*}\, ,
\nn\eea
and  matches the algebra~$\frak p$ of~\eqn{alg}with~$Q_\alpha=({\bf d,d^*})$,~${\cal D}=\bm\Delta$ and~$p=q=1$,~$r=0$.
We compute the BRST detour quantization of the rank~1, first class constraint algebra in the first line above in Section~\ref{formdetour}.


\subsection{$\mathfrak{so}(2,2)\!\supset\! \frak{gl}(2)$: K\"ahler Geometry}\label{kahlersusy}
The last model we would like to discuss is the spinning particle with~${\cal N}=4$ (in real counting) supersymmetries on the worldline, propagating on K\"ahler background (see~\cite{Marcus}~\cite{Fiorenzo}~\cite{Bellucci:2001ax} for more details).
We now denote by  ($M,g_{\mu\nu}$) a K\"ahler Manifold of complex dimension~$n$ with coordinates ~$x^{\mu}=(z^i,\bar z^{\bar \jmath})$
and Hermitean metric~$g_{i\bar\jmath}\ \! dz^i\otimes d\bar z^{\bar\jmath}$. 
Because of the special holonomy of $M$ the fermionic degrees of freedom split according to~$(\theta^\mu,\theta^{*\mu})=(\theta^i,\bar\theta^{\bar\imath}, \theta^{*i},\bar\theta^{*\bar\imath})$ and the first order action principle, which is the same as for the~${\cal N}=2$ model in the previous Section, can be simplified
$$
S^{(1)}=\int d\tau\Big[\dot x^\mu \pi_\mu + \frac i 2\  \!\theta^*_\mu \frac{\nabla\theta^\mu}{dt}
-\frac 12 \pi_\mu g^{\mu\nu}\pi_\nu+\frac 18 R^i{}_{\bar \jmath}{}^k{}_{\bar l} G_i^{\bar\jmath} G_k^{\bar l} \Big]\, ,
$$
where the $G_i^{\bar\jmath}\equiv\theta_i^* \bar \theta^{\bar\jmath}+\theta_i \bar \theta^{*\bar\jmath}$ are~$\frak u(n)$ Noether charges. This action is invariant under rigid
world line translations,~${\cal N}=4$ supersymmetry and~$\frak gl(2)$~$R$-symmetries.
Notice that the model realizes only a subgroup of  the maximal~${\cal N}=4$~$R$-symmetry algebra~$\frak{so}(2,2)\supset \frak{gl}(2)$.

The symplectic structure, operator identifications and Hilbert space of this model are, of course, identical to those of the previous Section for
the~$\frak{so}(1,1)$ model. In particular, wavefunctions are differential forms, but it is now possible to split them according to
the Hodge decomposition. However, since the background manifold~$M$ is K\"ahler, there are additional supersymmetry and~$R$-symmetry 
Noether charges
$$
Q=i\th^i \pi_i\, ,\qquad \bar Q = i\bar\th^{\bar \imath}\pi_{\bar \imath}\, ,
$$
$$
\bar Q^* = i\bar\th^{*\bar \imath}\pi_{\bar \imath}\, ,\qquad Q^*=i\th^{*i} \pi_i\, ,
$$
\vspace{-.3cm}
$$
G_i^{\bar\jmath}\equiv\theta_i^* \bar \theta^{\bar\jmath}+\theta_i \bar \theta^{*\bar\jmath}\, .
$$
At the quantum level, the Hamiltonian again corresponds to the form Laplacian while the supercharges 
become Dolbeault operators
$$
Q\to{\bm  \partial} ,\qquad \bar Q \to{ \bar{ \bm \partial}}\, ,
$$
$$
\bar Q^* \to \bar{\bm \delta}\qquad Q^*\to \bm \delta\, ,
$$
where~${\bf d}={ \bm \partial}+{ \bar{\bm \partial}}$ and~${\bf d^*}={\bm \delta}+{\bar{\bm \delta}}$.
The (non-trivial)~$R$-symmetries become the generators of the~$\frak{sl}(2)$
Lefschetz algebra ~$\{{ \bm \Lambda},{\bf N}, {\bf L}\}$ along with an additional~$\frak u(1)$ generator 
that computes the difference of holomorphic and antiholomorphic form degrees. Explicitly these are given by
$$
\begin{array}{rclrclrcl}
&&&{\bf N}&=&dz^i\frac{\partial}{\partial (dz^i)}+d\bar z^{\bar\imath}\frac{\partial}{\partial (d\bar z^{\bar\imath})}\, ,&&&\nonumber\\[4mm]
\bm \Lambda&=&\frac{\partial}{\partial(dz^i)}g^{\bar\jmath i}\frac{\partial}{\partial(d\bar z^{\bar\jmath})}\, ,&&&&\bf L&=&dz^i g_{i\bar\jmath}d\bar z^{\bar \jmath}\, ,\\[4mm]
&&& {\bf U}&=&dz^i\frac{\partial}{\partial (dz^i)}-d\bar z^{\bar\imath}\frac{\partial}{\partial (d\bar z^{\bar\imath})} \, .&&&
\end{array}
$$
The Lefschetz algebra acts on differential forms in a way that is rather analogous to the~$\frak{sp}(2)$ algebra
on symmetric forms in Section~\ref{sp2}, but where the K\"ahler form plays the role of the metric.

The differential symmetry generators  obey the algebra~$\frak p$ in~\eqn{alg} with~$p=q=2,r=0$, 
\be\label{Kalg1}
\{\bm \delta,\bm \partial\}=\frac12\,  \bm\Delta=\{\bar{\bm \delta},\bar{\bm \partial}\},
\ee
while~$(\bm \partial,\bar {\bm \delta})$ and~$(\bar{\bm \partial},-\bm\delta)$ form doublets under the~$\frak{sl}(2)$ Lefschetz algebra given by
\be\label{Kalg2}
[\N,\bm \Lambda]=-2\bm \Lambda\, ,\qquad [\bm \Lambda,{\bf L}]=n-\N\, ,\qquad [\N,{\bf L}]=2{\bf L}\, .
\ee
In the following, we will focus on the rank~1, first class constraint algebra~$\{\bm\Delta,\bm \partial,\bar{\bm \partial},\bm \delta,\bar{\bm \delta}\}$
along with the case where we additionally gauge the K\"ahler form operator~$\bf L$ (or equivalently ${\bm \Lambda}$). We have now gathered together
our quantum mechanical examples and are ready to study their BRST detour quantization.


\section{BRST Detour Cohomology}
\label{examples}
In this Section we use the general method presented in Section~\ref{BRST} to construct detour complexes of the first class constraint algebras corresponding to the quantum mechanical models described in Section~\ref{SUSY QM}. The first two examples have appeared in the literature before, but fit extremely neatly into our detour machinery (see~\cite{Bastianelli:2009eh} for an overview and original references). Readers interested only in the novel K\"ahler results can safely skip directly to Section~\ref{KD}.

\subsection{$\mathfrak{sp}(2)$ Spinning Particle: Higher Spins}\label{sp2BRST}
The first example we study is the BRST detour quantization of the~$\mathfrak{sp}(2)$ spinning particle model given in Section~\ref{sp2}. In the language of Section~\ref{BRST} this is the case~$p=q=0, \,r=1$ with the following ingredients
\begin{center}
\begin{tabular}{c|c}
\hline\\[-4mm]
Algebra~$\frak p$ &~$[{\bf div} ,{\bf grad}]=\bm\Delta\, $\\[2mm]\hline\\[-3mm]
Hilbert Space
~${\cal H}$ &~$\Gamma(\odot {T^* M })$\\[2mm]\hline\\[-3mm]
\begin{tabular}{c}BRST \\Hilbert Space~${\cal H}_{\rm BRST}$\end{tabular} & 
${\cal H}[[z , p]] \oplus c {\cal H}[[z , p]] $\\[4mm]\hline\\[-3mm]
BRST charge&
\begin{tabular}{c}
$Q_{\rm BRST}=c\bm\Delta+Q-M\frac{\partial}{\partial c}$\\[2mm]
$Q=\GRAD\frac{\partial}{\partial p} +z\DIV\, ,\quad$
$M= z \frac{\partial}{\partial p}$
\end{tabular}\\
\hline
\end{tabular}
\end{center}
This is a very simple example because elements of~${\cal H}_{\rm BRST}$ are terminating power series as all ghosts are Grassmann odd. Explicitly, an arbitrary element is of the form
\bea
\Psi_{\rm BRST} &=&\psi+c\chi \ ,\nn\\[4mm]
 \psi&=&\psi_{00} + p \psi_{10} + z \psi_{01} + pz\psi_{11}\,  ,\nn\\[2mm]
\chi&=&\chi_{00} + p \chi_{10} + z \chi_{01} + pz\chi_{11}\,  \, ,  \nn
\eea
where~$\psi_{ij},\chi_{ij}$ are symmetric forms on~$M$. 

There is but one ghost combination of the form~\eqn{ys}:
\be  \nn
y=2pz \, .
\ee
(Importantly, although $y$ is even, it obeys $y^2=0$ thanks to its origin as a product of ghosts.)
Thus we readily compute\footnote{Of course $\frac y2{\cal H}$ and $y {\cal H}$ are equivalent spaces, 
but we keep the factor $\frac12$ to denote the choice of basis $(X_{00},X_{11})=X_{00} + \frac y 2 X_{11}$. }
\bea
\ker M\ \ &=&{\cal H} \oplus z{\cal H} \oplus  \frac y2{\cal H}\, , \nn \\[2mm]
\ker M^{\mathfrak s}\ &=&{\cal H} \oplus p{\cal H} \oplus \frac y2 {\cal H}\, ,  \nn\\[2mm]
\ker M \cap \ker M^{\mathfrak s}&=& {\cal H} \oplus  \frac y2 {\cal H}\, , \nn\\[2mm]
{\rm im} M\ \  &=&{\rm im} MM^{\mathfrak s}\ = \ z {\cal H}  \nn\, .
\eea
This implies that $M^{-1}$ is defined by
\bea \nn
M^{-1}:{\rm im} M &\to& {\rm im} M^{\mathfrak s} \, ,
\\[2mm]
M^{-1}(z\alpha)&=&p\alpha \, ,
\eea
for all~$\alpha\in\cal{H}$.

The incoming differential~$Q_{{\rm ker} M^{\mathfrak s}}:p{\cal H} \to {\cal H}\oplus \frac y2{\cal H}$ acts on ghost number~$-1$ fields~$\psi_{10}$ as left multiplication by the row matrix
\be \label{insp2}
Q_{{\rm ker} M^{\mathfrak s}}=
\left(\begin{array}{c}\GRAD\\[2mm]-\DIV\end{array}\right) \, ,
\ee
and acts trivially at ghost number $1$ giving the incoming complex
\be \label{sp2incoming}
\mbox{
$
0\longrightarrow p{\cal H} \stackrel{\tiny\left(\!\!
\begin{array}{c}
\GRAD
\\[2mm]-\DIV
\end{array}\!\!\right)}{\longrightarrow}
{\cal H}\oplus \frac y2{\cal H}
\stackrel{\tiny }{ \longrightarrow }
0$\, .} 
\ee
Similarly, the outgoing differential $Q: {\cal H}\oplus \frac y2{\cal H}\to z{\cal H}$
acts on the independent fields~$\chi_{\ker M}=\chi_{00}+\frac12 y\chi_{11}$ as left multiplication by the row matrix
\be \label{outsp2}
\left(\begin{array}{cc}\DIV & \GRAD \end{array}\right) \, 
\ee
giving the outgoing complex
\be \label{sp2outgong}
\mbox{
$0\longrightarrow 
{\cal H}\oplus \frac y2{\cal H}
\stackrel{\tiny
\left(\begin{array}{cc}\!\!\!\!\DIV \ \GRAD \!\!\!\!\end{array}\right)
}{-\!\!\!-\!\!\!\longrightarrow}
z{\cal H} 
\stackrel{\tiny }{ \longrightarrow }
0$\, .} 
\ee
The long operator $(\bm\Delta - Q M^{-1} Q) : {\cal H} \oplus \frac{y}{2}{\cal H} \to {\cal H} \oplus \frac{y}{2}{\cal H}$ acts on the ghost number zero fields~$\psi_{00}+\frac 1 2 y\psi_{11}$ as left multiplication by the square matrix
\be
\left(
\begin{array}{cc}
\bm
\Delta-\GRAD\ \DIV &-\GRAD^2 
\\[2mm]  \DIV^2 & \bm\Delta+\DIV\GRAD
\end{array}
\right)
\label{detoursp2}
\ee
which connects the incoming and outgoing complexes~\eqn{sp2incoming},~\eqn{sp2incoming} to form a detour complex 
\be
0\longrightarrow {\cal H} \stackrel{\tiny\left(\!\!
\begin{array}{c}
\GRAD
\\[2mm]-\DIV
\end{array}\!\!\right)}{\longrightarrow}
{\cal H}^2 
\stackrel{\tiny \left(\!\!
\begin{array}{cc}
\!\bm\Delta-\GRAD\DIV\! 
& -\!\GRAD^2\! 
\\[2mm]  \!\-\DIV^2\! 
& \!\bm\Delta+\DIV\GRAD\!
\end{array}
\!\!\right)}{-\!\!\!-\!\!\!-\!\!\!-\!\!\!\longrightarrow }
{\cal H}^2 
\stackrel{{}^{{}^{\tiny(\!\!\!\!\begin{array}{c}\DIV\ \GRAD\end{array}\!\!\!\!\!)}}}{\longrightarrow}{\cal H}\to 0\, ,
\ee
of the form~\eqn{detcomplexfull} which encodes two equations of motion for the physical (ghost number zero) fields~$(\psi_{00},\psi_{11})$
as well as their gauge invariances and Bianchi identities in terms of the incoming and outgoing complexes, respectively.

A simpler system 
can be obtained by gauging the~$R$-symmetry corresponding to $y$
using the method described in~Section~\ref{R symmetry}. 
This $R$-symmetry is determined by 
$[Q,y]=[Q,\rho] $, which implies $\rho= \G$
(see Section~\ref{sp2}).
We define
\bea \nn
T=\G-y \, , & &  \, T^{\frak *}=\TR-y^{\frak *}\, ,\\[2mm]
y^{\frak *}&=&2\frac{\partial}{\partial p}\frac{\partial }{\partial z}\,  , \nn
\eea
and note that $y^*$ is the ghost bilinear associated to the $R$-symmetry $\TR$;
\be \label{T*QM}
[T^*,Q]=[T^*,M]=0 \, .
\ee
These commutation relations imply that $Q, G: {\rm ker} \, T^*\to {\rm ker}\, T^*$ and for this reason
we proceed to describe the sector 0  ($={\rm coker} T $) 
detour complex~\eqn{kerRBcomplex} by choosing ${\rm coker} \,T $ representatives 
in~${\rm ker} \,T^{\frak *}$; all boundary maps in the detour complex preserve this choice.

With this choice, elements of the detour complexes~\eqn{kerRBcomplex} and~\eqn{kerR*Bcomplex} satisfy
the conditions tabulated below
\begin{center}
\begin{tabular}{|c|c|c|c|}  
\hline
~~\raisebox{-3mm}{$N_{\rm gh}$}~~&~~~\raisebox{-3mm}{Fields}~~~&~~$\ker T^{\frak *}$ ~~ &~~~$\ker T$ ~~~ \\[-2mm]
&&conditions&conditions\\
\hline &&&
\\[-3mm]
-1&$p\psi_{10}$&$\TR \psi_{10}=0$&$\G \psi_{10}=0$
\\[3mm]
0&$\psi_{00}+y\psi_{11}$&\!$\psi_{11}=-\frac12\TR \psi_{00}, \, \TR \psi_{11}=0~$\!&$\psi_{00}=\frac12\G \psi_{11}, \G\psi_{00}=0$\\[3mm]
1&$z\chi_{01}$&$\TR \chi_{01}=0$&$\G \chi_{01}=0$\\[2mm]
\hline
\end{tabular}
\end{center}
Using $\ker \G=0$ we quickly determine that ${\rm ker} \, T$ is trivial, 
so the sector~1 detour complex is empty. 
In the sector~0 detour complex is then given by

\bea \label{2compdet}
\  {\ss 0}\to {\ss \ker \TR} \!\!\!\stackrel{\tiny\left(\!\!\!
\begin{array}{c}
\!\GRAD\!
\\[2mm] \!\frac12\!\DIV\TR\!
\end{array}\!\!\!\right)}{\longrightarrow}
\!\!\!
{\ss (1 -\frac14y\TR){\ker \TR^2}} 
\hspace{-.9cm}
\stackrel{{}^{{}^{
\tiny \left(\!\!
\begin{array}{cc}
\!\!\bm\Delta\!\!-\!\!\GRAD\DIV\!\! 
& \!\!\frac12\!\GRAD^2\! \TR\! \!
\\[2mm]  \!\!-\DIV^2\!\!
& \!\!-\!\frac12\! \bm\Delta\!\TR\!\!+\!\!\frac12\!\!\DIV\!\GRAD\!\TR\!\!
\end{array}
\!\!\!\right)
}}
}
{-\!\!\!-\!\!\!-\!\!\!-\!\!\!\longrightarrow }
\hspace{-.9cm}
{\ss (1-\frac14y\TR){\ker \TR^2} }\!\!\!
\stackrel{{}^{{}^{\tiny(\!\!\!\!\begin{array}{c}\DIV\ \GRAD\end{array}\!\!\!\!\!)}}}{\longrightarrow}
\!\!\!{\ss \ker \TR}\to {\ss 0}\, .\nn\\
\eea
There is a redundancy in this detour complex: the equation of motion for~$\psi_{00}$ in the first component implies that of the second. The same is true of the gauge symmetry and Bianchi identity. Thus, all of the cohomology information is contained in the single component detour complex
\be \label{singlecompdet}
0\longrightarrow {\ker \TR} \stackrel{\GRAD}{\longrightarrow}
\ker \TR^2 
\stackrel{(\bm\Delta -\GRAD \DIV -\frac 1 2\GRAD^2 \TR) }{-\!\!\!-\!\!\!-\!\!\!-\!\!\!\longrightarrow }
\ker \TR^2 
\stackrel{\DIV}{\longrightarrow} \ker \TR\to 0\, .
\ee
However, the detour operator $$G_{\rm sing}\equiv \Delta -\GRAD \DIV -\frac 1 2\GRAD^2 \TR$$ of this single component complex is not self adjoint. It is possible to obtain a self adjoint version of the detour operator by the  ghostbusting procedure discussed in Section~\ref{R symmetry}, which essentially allows us to
trade the dependence on $y$ for the operator $\G$.
Explicitly the long operator of the two component detour complex~\eqn{2compdet} acts on the single field $\psi_{00}\in {\cal H}$ as
\bea \nn
G=\Delta-\GRAD\DIV+\frac12 \GRAD^2\TR ~~~~~~~~~~~~~~~~~\nn\\
\nn
~~~~~~~~~~~~~~~~~~~~~~~+\frac12 y(\DIV^2 -\frac12 \G \Delta\TR -\frac12 \DIV\GRAD \TR) \, .
\eea
Ghostbusting and arranging factors by form degree yields the manifestly self adjoint ``Einstein operator"
\bea \nn
{\cal G}=\Delta-\GRAD\DIV+\frac12 \GRAD^2\TR ~~~~~~~~~~~~~~~~~~~\\
~~~~~~~~~~~~~~~~~~~~~~~+\frac12 \G\DIV^2 -\frac12 \G \Delta\TR -\frac14 \G \GRAD\DIV \TR \, . \nn
\eea

It is also convenient to note that, since the long operator preserves $\ker T^*$, fields in its image are of the form
$(1-\frac14 y\TR)\phi$. Combining this observation with the process of ghostbusting one obtains the alternative expression 
\bea
{\cal G}=(1-\frac14 \G\TR)(\Delta-\GRAD\DIV+\frac12 \GRAD^2\TR)\, 
\eea
in which the gauge invariance $\delta\psi_{00}=\GRAD\a $ for $\a\in \ker \TR$ is manifest.
We note that the self adjoint version of the long operator is obtained by left multiplication of the invertible factor $1-\frac14 \G\TR$ on the single component long operator $G_{\rm sing}$ of~\eqn{singlecompdet}.

Finally let us discuss the physics underlying this complex. 
When~$\psi_{00}$ is a 1-form, the ghostbusted detour equation of motion $ {\cal G}\psi_{00}=0$ amounts precisely to Maxwell's equations. For a symmetric two-form,
${\cal G}\psi_{00}=0$ is exactly the linearized Einstein tensor---hence its name. 
When~$\psi_{00}$ is an arbitrary rank~$s$ symmetric tensor the result is the equations of motion for a spin~$s$, massless field
as first described by Curtright and Fronsdal~\cite{Fronsdal:1978rb,Curtright:1979uz}. For an introduction to the theory of higher spins, the  review articles~\cite{Vasiliev:2004qz}) are rather useful.

\subsection{$\mathfrak{so}(1,1)$ Spinning Particle: ~$p$-Form Electro\-magnetism}\label{formdetour}
The next example is the BRST quantization of the~${\cal N}=2$ supersymmetric quantum mechanical model given in Section~\ref{so11}. 
The following computation first appeared in\footnote{We thank Boris Pioline and Andy Neitzke for a partial collaboration in that work.}~\cite{Bastianelli:2009eh}.
By now the recipe for constructing the BRST operator and Hilbert space from the underlying quantum mechanical one should be clear so we tabulate the main ingredients

\begin{center}
\begin{tabular}{c|c}
\hline\\[-4mm]
Algebra~$\frak p$ &~$\{ {\bf d,d^*} \}=\bm\Delta$\\[2mm]\hline\\[-3mm]
Hilbert Space~${\cal H}$ &~$\Gamma(\wedge {M})$\\[2mm]\hline\\[-3mm]
\begin{tabular}{c}BRST \\Hilbert Space~${\cal H}_{\rm BRST}$\end{tabular} & 
${\cal H}[[z , p]] \oplus c {\cal H}[[z , p]]\ni \sum\frac{z^sp^t}{s!t!}\Big(\psi_{st}+c\chi_{st}\Big)$\\[4mm]\hline\\[-3mm]
BRST charge&
\begin{tabular}{c}
$Q_{\rm BRST}=c\bm\Delta+Q-M\frac{\partial}{\partial c}$\\[2mm]
$Q={\bf d}\frac{\partial}{\partial p} +z{\bf d^*}\, ,\quad$
$M= z \frac{\partial}{\partial p}$
\end{tabular}\\
\hline
\end{tabular}
\end{center}
So far, the main contrast with the~$\frak{so}(1,1)$ example of the previous Section
is that the ghost and antighost~$(z,p)$ are {\it bosonic} so BRST wavefunctions
can be expanded in them to arbitrary order.

Our next task is to compute~$\ker M$ and fix a gauge for~$\coker  M$, to which end we introduce the 
adjoint operation  of~\eqn{sadj} \be \label{formsadj}
z^{\mathfrak s} = \frac{\partial}{\partial z}\, ,\qquad p^{\mathfrak s} = \frac{\partial}{\partial p}\, .
\ee
In particular, we notice that  {\it no} ghost combinations of the form~\eqn{ys} exist. 
With this information we readily compute
\bea
\ker M\ \ &=& {\cal H}[[z]] \, , \nn \\[2mm]
\ker M^{\mathfrak s}\ &=& {\cal H}[[p]] \, ,  \nn \\[2mm]
\ker M \cap \ker M^{\mathfrak s}&=& {\cal H} \, , \nn \\[2mm]
{\rm im} M\ \  &=&{\rm im} MM^{\mathfrak s}\ = \ z {\cal H}[[z,p]] \, . \nn 
\eea
This means that the partial inverse 
$
M^{-1}:{\rm im} M\to {\rm im} M^{\mathfrak s}
$
acts on~$\ker M$ at ghost number one as
\be \nn 
M^{-1}(z\alpha)=p\alpha\, ,
\ee
for any differential form~$\alpha\in {\cal H}$.

The incoming complex of this model consists of fields of the form
\be \nn 
\psi_{\ker M^\frak{s}}=\sum_{t>0}\frac {p^t}{t!}\psi_{0t} \in {\rm ker} M^{\mathfrak s}\, ,
\ee 
with differential 
$${Q}_{\ker M^\frak{s}} ={\bf d}\frac{\partial}{\partial p}\, .$$
Grading by ghost number  reproduces the de Rham complex 
\be \nn 
{\bf d} \psi_{0t}=0
 \qquad\qquad \psi_{0t}\sim\psi_{0t}+{\bf d}\alpha_{0t+1}~~~~~~~t> 0.
 \ee
In pictures we have
\be \nn
\shabox{
$\cdots\stackrel{\bf d}\longrightarrow \Gamma(\Lambda { M}) \stackrel{\bf d}\longrightarrow \Gamma(\Lambda { M}) \longrightarrow 0$}
\ee

Once we impose~$Q_{\ker M^\frak{s}}$~$\psi_{\ker M^\frak{s}}$~$=$~$0$, it follows that 
$Q$~$\psi_{\ker M^\frak{s}}$~$\in$~${\rm im} M$ and, therefore, that~$M^{-1}$ is well-defined on~$Q$~$\psi_{\ker M^\frak{s}}$.
At ghost number~0 we compute the action of the long operator 
\be \nn 
(\bm\Delta - Q M^{-1} Q) : {\cal H} \to {\cal H} 
\ee
on~$\left.\left(\psi_{\ker M^\frak{s}}\right) \right|_{N_{\rm gh}=0}\equiv\psi_{00}$ and find the closure condition 
\be  \nn 
\shabox{
${\bf d^*d}\psi_{00}=0\, .$
}
\ee
Lastly we calculate the outgoing complex, whose modules are the ghost number eigenspaces of~$\ker M={\cal H}[[z]]$. On this space 
$Q={\bf d}\frac{\partial}{\partial p} + z{\bf d^*}=z{\bf d^*}$, and we obtain the dual de Rham complex

\be \nn 
\shabox{
$0\longrightarrow \Gamma(\Lambda { M}) \stackrel{\bf d^*}\longrightarrow \Gamma(\Lambda { M}) \stackrel{\bf d^*}\longrightarrow \cdots$}
\ee
We  now connect the incoming and outgoing complex by the methods of Section~\ref{detour} and find the Maxwell detour complex
\be\nn
\shabox{$
\begin{array}{c}
\cdots
\stackrel{\bf   d}{\longrightarrow}
{\cal H}
\stackrel {\bf d }{\longrightarrow} 
{\cal H}
\stackrel{\bf  d} {\longrightarrow}
{\cal H}
\qquad
\qquad
\qquad
{\cal H}
\stackrel{\bf  d^*}{\longrightarrow} 
{\cal H}
\stackrel{ \bf d^*}{\longrightarrow}
{\cal H}
\stackrel{\bf d^*}{\longrightarrow}
\cdots
\\
\ \Big|\hspace{-1mm}\raisebox{-2.5mm}{\underline{ \ \quad\ \ \ \raisebox{1mm}{$ {\bf d^*  d}$ }\quad \ \ \  \ \ }} \hspace{-1.3mm}{\Big\uparrow}
\end{array}
$}
\ee
Notice that the ghost number zero cohomology~$\ker\!\left({\bf d^* d}\right)\! / {\rm im} ({\bf d})$ 
is exactly the solution space to the equations of motion of~$p$-form electromagnetism.

\subsection{$\mathfrak{so}(2,2)\!\supset\! \frak{gl}(2)$ Spinning Particle:~$(p,q)$-Form Electromagnetism}
\label{KD}
Our last example of a physical model with generalized supersymmetry algebra of the form~\eqn{alg}
is the~${\cal N}=4$ spinning particle on a K\"ahler manifold. The underlying supersymmetric quantum mechanical model was described in Section~\ref{kahlersusy} of this Article;
the main formulae were originally announced  in~\cite{Cherney:2009vg}. 
This model is the first with both an infinite tower of supersymmetry ghosts {\it and} a non-trivial kernel of the operator~$M$.
We begin by tabulating the main BRST kinematical data:

\begin{center}
\begin{tabular}{c|c}
\hline\\[-4mm]
Algebra~$\frak p$ &~$\{ \bm \partial,\bm \delta \}=\frac12 \bm\Delta=\{\bar{\bm \partial},\bar{\bm \delta}\}$\\[2mm]\hline\\[-3mm]
Hilbert Space~${\cal H}$ &~$\Gamma(\wedge {M})$\\[2mm]\hline\\[-3mm]
\begin{tabular}{c}BRST \\Hilbert Space~${\cal H}_{\rm BRST}$\end{tabular} & 
${\cal H}[[z,\bar z , p,\bar p]] \oplus c {\cal H}[[z,\bar z , p,\bar p]]$\\[4mm]\hline\\[-3mm]
BRST charge&
\begin{tabular}{c}
$Q_{\rm BRST}=\frac12\ c\bm\Delta+Q-M\frac{\partial}{\partial c}$\\[2mm]
$Q=z\bm \delta+\bar z\bar{\bm \delta}+\bm \partial\frac{\partial}{\partial p}+\bar{\bm \partial}\frac{\partial}{\partial \bar p} \, ,\quad$
$M=z\frac{\partial}{\partial p}+\bar z\frac{\partial}{\partial \bar p}~$
\end{tabular}\\[5mm]
\hline
\end{tabular}
\end{center}
 There is a single ghost bilinear of the form~\eqn{ys} which we name
\be \nn 
y=p\bar z-z \bar p \, .
\ee
In terms of this variable the incoming and outgoing modules and domain of the long operator are, respectively
\bea 
\ker M^{\mathfrak s}\ &=&{\cal H}[[y,p,\bar{p}]]\, ,\nn 
\\[2mm]
\ker M\ \ &=&{\cal H}[[z,\bar{z},y]]\, ,\nn 
\\[2mm]
\ker M \cap \ker M^{\mathfrak s}&=& {\cal H}[[y]]\, .\nn 
\eea
The incoming complex 
has differential~${Q}_{\ker M^\frak{s}}: \ker M^{\mathfrak s}\rightarrow \ker M^{\mathfrak s}$, determined  by first acting with the full operator~$Q$ on~$\psi_{\ker M^\frak{s}}$ and then 
finding the unique representative~${Q}_{\ker M^\frak{s}}\psi_{\ker M^\frak{s}}$ of the ${\rm coker}\, M$ class 
$[Q\psi_{\ker M^\frak{s}}]$ in $\ker M^{\frak s}$. The result is the operator
\be
\label{so22 incoming op}
{Q}_{\ker M^\frak{s}}
=\left(1+\frac{y\frac{\partial}{\partial y}}{2-N_{\rm gh}}\right)\left[{\bm \partial}\frac{\partial}{\partial p}
+\bar{\bm \partial } \frac{\partial}{\partial \bar p}\right]\ 
+\ \frac{y}{2-N_{\rm gh}}\ \left[ \bar {\bm \delta}\frac{\partial}{\partial p} 
-  \bm \delta\frac{\partial}{\partial \bar p} \right] \, . 
\ee
Importantly, note that this operator is defined on ${\cal H}[[p,\bar{p},y]]$ so that~$\partial y/\partial p =\partial y/\partial \bar p\equiv 0$.
A useful check of this result is that~${Q}_{\ker M^\frak{s}}$ really squares to zero and therefore determines the incoming complex~\eqn{kerIncoming}.

On the outgoing complex, the differential $Q:\ker M\to \ker M$ acts as
\be \label{so22 outgoing op}
Q|_{\ker M}=z\bm \delta +\bar{z}\bar{\bm \delta} + (\bar{z}{\bm \partial} -z\bar{\bm \partial} )\frac{\partial}{\partial y} \, .
\ee
The detour operator ~${G}=\frac12\bm\Delta-QM^{-1}Q$  which connects the incoming and outgoing complexes is
\be \label{so22 long op}
\shabox{
${G}=(\frac 12 {\bm\Delta}-\bm \partial \bm \delta- \bar{\bm \partial}\bar{\bm \delta})(1+y\frac{\partial}{\partial y}) 
- \bar{\bm \partial}\bm \partial(y\frac{\partial^2}{\partial y^2}+2\frac{\partial}{\partial y})
-\bar{\bm \delta}\bm \delta y$
}
\ee
The above results are neatly summarized in a single pictograph :
\begin{table}[h]\label{kahdetour}
{\footnotesize
$$
\begin{CD}
@>{\begin{array}{c}\scriptscriptstyle\left(1+\frac{y\frac{\partial}{\partial y}}{2-N_{\rm gh}}\right)\left[{\bm \partial}\frac{\partial}{\partial p}+\bar{\bm \partial } \frac{\partial}{\partial \bar p}\right]\\
\scriptscriptstyle+\frac{y}{2-N_{\rm gh}}\ \left[ \bar {\bm \delta}\frac{\partial}{\partial p}-  \bm \delta\frac{\partial}{\partial \bar p}\right]\end{array}}>>
\!\!\!{\cal H}[[p,\bar p,y]]\!\!\! 
@>{\begin{array}{c}\scriptscriptstyle\left(1+\frac{y\frac{\partial}{\partial y}}{2-N_{\rm gh}}\right)\left[{\bm \partial}\frac{\partial}{\partial p}+\bar{\bm \partial } \frac{\partial}{\partial \bar p}\right]\\
\scriptstyle+\frac{y}{2-N_{\rm gh}}\ \left[ \bar {\bm \delta}\frac{\partial}{\partial p}-  \bm \delta\frac{\partial}{\partial \bar p}\right]\end{array}}>> \!\!\cdots\!\!\!\! @>\begin{array}{c}\scriptscriptstyle\left(1+\frac13y\frac{\partial}{\partial y}\right)\left[{\bm \partial}\frac{\partial}{\partial p}+\bar{\bm \partial } \frac{\partial}{\partial \bar p}\right]\\ +\scriptstyle\frac{y}{3} \left[ \bar {\bm \delta}\frac{\partial}{\partial p}-  \bm \delta\frac{\partial}{\partial \bar p}\right]\end{array}>> \!\!{\cal H}[[y]]   \\
@. @. @. \hspace{-.01mm}\Big|\\[-1mm]
@. @. @. \hspace{-.01mm}\Big|\\[-1mm]
@. @. @.  
@V(\frac 12 {\bm\Delta}-\bm \partial \bm \delta- \bar{\bm \partial}\bar{\bm \delta})(1+y\frac{\partial}{\partial y}) 
- \bar{\bm \partial}\bm \partial(y\frac{\partial^2}{\partial y^2}+2\frac{\partial}{\partial y})
-\bar{\bm \delta}\bm \delta yVV\\[-1mm]
@.@.@.\hspace{-.01mm}\Big|\\[-1mm]
@.@.@.\hspace{-.01mm}\Big|\\
@<\phantom{spa}z\bm \delta+\bar z \bar{\bm \delta}+(\bar z \bar{\bm \delta}-z\bar{\bm \partial})\frac{\partial}{\partial y}\phantom{spa}<< 
\!\!\!{\cal H}[[z,\bar z, y]]\!\!\! 
@<\phantom{spa}z\bm \delta+\bar z \bar{\bm \delta}+(\bar z \bar{\bm \delta}-z\bar{\bm \partial})\frac{\partial}{\partial y}\phantom{spa}<< \!\!\!\cdots\!\! 
@<\phantom{spa}z\bm \delta+\bar z \bar{\bm \delta}+(\bar z \bar{\bm \delta}-z\bar{\bm \partial})\frac{\partial}{\partial y}\phantom{spa}<< \!\!{\cal H}[[y]] 
\end{CD}
$$
}
\end{table}

This complex is perhaps not the one of maximal physical interest because the ghost number zero cohomology still involves the ghosts through
the variable~$y$. Indeed, Taylor expanding in~$y$ gives a complex involving an infinite tower of fields at ghost number zero.
By gauging an~$R$-symmetry we can remove the~$y$ dependence, but first we note that there is an interesting structure underlying
the long operator~$G$. To begin with, rather than working with an infinite tower of fields, we can view the bosonic variable~$y$ as an 
additional coordinate on the manifold~$\Real\times M$. Moreover, notice that the triplet of operators appearing in the long operator~$G$
\be \label{ghost alg}
e=y\, ,\qquad h=y\frac{\partial}{\partial y}+1\, ,\qquad f=y\frac{\partial^2}{\partial y^2}+2\frac{\partial}{\partial y}\, ,
\ee
generate the~$\frak{so}(2,1)\cong \frak{sl}(2)$ Lie algebra
\be\nn
[e,f]=-2h\, ,\qquad [h,e]=e\, ,\qquad [h,f]=-f\, .
\ee
(In fact, Fourier transforming~$y\leftrightarrow -i \partial/\partial y$, this is the infinitesimal action of the one-dimensional conformal 
group~${\rm conf}(\Real)$.) In addition, allowing ourselves to formally invert $\bm\Delta$, the triplet of operators defined by
\be \nn
\bm\Delta F=2\bm \delta \bar{\bm \delta}\, , \qquad -2\bm\Delta H=\bm\Delta -2\bm \partial \bm \delta -2\bar{\bm \partial} \bar{\bm \delta}\, ,\qquad
\bm\Delta E=2\bm \partial \bar{\bm \partial}\, ,
\ee
also obey the the~$\frak{sl}(2)$ Lie algebra
\be\nn
[E,F]=-2H\, ,\qquad [H,E]=E\, ,\qquad [H,F]=-F\, .
\ee
Therefore we obtain a simple result for the long operator that is  manifestly invariant under the  Lefschetz~$\frak{sl}(2)$ algebra
\be\nn
G= Ef+Hh+Fe\, .
\ee
This elegant, geometric result can be extended to models with higher~$R$-symmetries, as discussed in~\cite{CLW}.
However, rather than pursuing this direction, we now remove the ghosts remaining at ghost number zero  by gauging
an~$R$-symmetry.

Following the method prescribed in Section~\ref{R symmetry} we gauge the $R$-symmetry corresponding to~$y$. Requiring~$[Q,y]=[Q,\rho]$ implies $ \rho=\L$,
where $\bm L$ is the K\"ahler form  operator discussed in Section~\ref{kahlersusy}. The operators $T$ and $\bar T^*$ are given by
\bea\nn
T=\L-y ~ , ~~~~~
\bar{T}^*=\bm \Lambda-\bar{y}^* \, ,
\\[2mm] \label{y^*}
\bar{y}^* = \Big(
{\frac{\partial}{\partial p}\frac{\partial}{\partial \bar z}-\frac{\partial}{\partial \bar p}\frac{\partial}{\partial z}}
\Big)\, ,~~~~~
\eea
by virtue of $\La \equiv \bar{\L}^*$.
The kernel of $T$ is  again empty 
in $\ker M \cap \ker M^{\frak s}={\cal H}[[y]]$ 
since, expanding in powers of $y$,
\bea
(\L-y)\sum_{n} y^n \psi_{n}=0 \implies  \psi_{m}=\L^{\!n} \, \psi_{m+n} ~\, \forall \,m,n\in\mathbb{N} \, ,
\eea
and $\L^{\frac{\dim M}{2} +1}=0$ on any finite dimensional K\"ahler manifold $M$. Thus sector~1 is empty.

Sector~0 consists of solutions to the 
operator equation
\be\nn
\bar{T}^*\Psi=(\bm \Lambda-\bar{y}^*)\Psi=0 \, .
\ee
On $\ker M={\cal H}[[z,y]]$ or ${\cal H}[[p,y]]$ this condition may be expressed as a Bessel-type equation
\bea \label{BesselDE}
0&=&\Big[\bm \Lambda-(2+N)\frac{\partial}{\partial y}-y\frac{\partial^2}{\partial y^2}\Big]\psi\, ,
\eea
where
\bea\nn
N&\equiv &
p\frac{\partial }{\partial p} +\bar{p}\frac{\partial }{\partial \bar{p}}
+z\frac{\partial }{\partial z}+ \bar{z} \frac{\partial }{\partial {z} }\, .
\eea
Expanding~$\psi(y)=\sum_{n}y^n \psi_n$ at a definite value of $N$, one finds
\be
\psi_{n}=\frac{N!}{(N+n+1)!n!} \La\!^n  \, \psi_0\, ,
\ee
 so that all coefficients~$\psi_{n>0}$ are determined by an arbitrary differential form~$\psi_0$. 

The same conclusion can be reached in the  $N_{\rm gh}=0$  case 
(which is also the $N=0$ case) by viewing~\eqn{BesselDE} as a differential equation in the variable~$y$ with $\La$ viewed as a constant.
There are two solutions to this equation but only the modified Bessel function of the second kind obeys the correct boundary conditions
(namely a polynomial expansion in~$y \La$) 
\bea \label{Bessel soln}
\psi(y)=\frac{I_1(2\sqrt{y\bm \Lambda})}{\sqrt{y\bm \Lambda}}\, \psi_0
\sim\Big(1+\frac12 y\bm \Lambda + \frac1{12} y^2 \bm \Lambda^2 + \frac 1{144} y^3\bm \Lambda^3+\cdots\Big)\, \psi_0 \, .
\eea
This solution agrees with the recurrence relation above. 

The commutation relations
\be
[Q,\bar{T}^*]=[M,\bar{T}^*]=0\, ,
\ee
guarantee that the long operator $G$ preserves $\ker \bar{T}^*$, and thus the power series 
form~\eqn{Bessel soln}. Therefore, it suffices to calculate the zeroth order of $G\psi$ since the entire expression is obtained by simply multiplying this term by the Bessel function in~\eqn{Bessel soln}.
We make use of~\eqn{so22 long op}  and~\eqn{ghost alg} to write
\be \nn
{G}=(\frac 12 {\bm\Delta}-\bm \partial \bm \delta- \bar{\bm \partial}\bar{\bm \delta}) 
(\bar{y}^*y - y \bar{y}^*)
- \bar{\bm \partial}\bm \partial \bar{y}^*
-\bar{\bm \delta}\bm \delta y\, ,
\ee
and from this, using  the~$\ker \bar{T}^*$ condition $\bar{y}^*=\La$,  quickly calculate the zeroth order term
\be \label{ez eom}
G_{{\rm sing}} \psi_{0} 
\equiv 
G\psi|_{y=0} = \La {\bm \partial} \bar {\bm \partial}\,  \psi_0 \, ,
\ee
and in turn  the action of the full long operator
\be \label{ghostly long op}
G\psi=\frac{I_1(2\sqrt{y\bm \Lambda})}{\sqrt{y\bm \Lambda}}\,  \La {\bm \partial} \bar {\bm \partial} \, \psi_0 \, .
\ee

The gauge and gauge for gauge symmetries, determined by the image of the incoming differential~\eqn{so22 incoming op} 
 at zeroth order in $y$, are then encoded by the incoming complex 
 \be
 \cdots \stackrel{\bm \partial \frac{\partial}{\partial p} + \bar{\bm \partial} \frac{\partial}{\partial \bar p}}{-\!\!\!-\!\!\!-\!\!\!-\!\!\!\longrightarrow}
 {\cal H}[[p,\bar p]] 
 \stackrel{\bm \partial \frac{\partial}{\partial p} + \bar{\bm \partial} \frac{\partial}{\partial \bar p}}{-\!\!\!-\!\!\!-\!\!\!-\!\!\!\longrightarrow}\cdots
 \ee 
Indeed the long operator is manifestly invariant under the gauge transformations 
\be\nn
\varphi\sim \varphi + \bm \partial \alpha + \bar{\bm \partial}\bar \alpha\, .
\ee

The Bianchi and``Bianchi for Bianchi" identities, determined by the image of the outgoing differential~\eqn{so22 outgoing op} 
 at zeroth order in $y$, are somewhat less obvious. Also, the long operator above is not  self adjoint. These problems are solved simultaneously  by the ghostbusting method described in~Section~\ref{R symmetry}. 
The ghostbusted representatives of the image of the long operator are obtained by replacing all factors of $y$ appearing (at the far left of each term) in~$G\psi$ with~$\L$. If we use the expression for $G$ appearing in~\eqn{so22 long op} and act on an arbitrary element 
\bea \nn
\Psi=\sum_{n=0}^\infty \frac{(y\La)^n}{n!(n+1)!} \ \psi_0
\eea
of sector 0 at ghost number zero, we obtain the manifestly self adjoint ghostbusted long operator~${\cal G}:{\cal H} \to {\cal H}$ 
\be\nn
{\cal G}=\  \scalebox{1.3}{:} \, {I_{0}(2\sqrt{\bm L\Lambda})}\, (\bm\Delta - 2\bm \partial\bm \delta-2\bar{\bm \partial}\bar{\bm \delta})
+\ 2\, \frac{I_{1}(2\sqrt{\bf L\Lambda})}{\sqrt{\bf L\Lambda}} \, (\bm \Lambda \bm \partial\bar{\bm \partial} +\bm \delta\bar{\bm \delta}{\bf L})
\, \scalebox{1.3}{:}\, ,\ \ \ee
where $:\bullet:$ indicates normal ordering by form degree.
Alternatively, we use the expression for $G$ appearing in~\eqn{ghostly long op} to obtain the alternative compact expression
\be \label{gb long op}
{\cal G} =  \scalebox{1.3}{:} \frac{I_1(2\sqrt{\bm L\bm \Lambda})}{\sqrt{\bm L\bm \Lambda}} \scalebox{1.3}{:} \,  \ \bm \Lambda {\bm \partial} \bar {\bm \partial}  \, .
\ee
In this form the long operator is manifestly gauge invariant
\bea \nn
{\cal G}\,{\bm \partial}={\cal G} \, \bar{\bm \partial}=0 \, 
\eea
The adjoint of~\eqn{gb long op}
\be \label{gb long po}
{\cal G}^*={\cal G} =  \   {\bm \delta} \bar {\bm \delta}  \L\ 
\scalebox{1.3}{:} \frac{I_1(2\sqrt{\bm L\bm \Lambda})}{\sqrt{\bm  L\bm \Lambda}}\scalebox{1.3}{:} \, ,
\ee
manifestly  satisfies the Bianchi identities
\bea \nn
\bm \delta \, {\cal G}=\bar{\bm \delta } \, {\cal G} =0\, 
\eea
These are precisely the Bianchi identities encoded by the outgoing operator~\eqn{so22 outgoing op} if one chooses the ghostbusted ${\rm coker}\ \!  T$ representatives for the outgoing complex.

The self adjoint operator ${\cal G}$ connects Dolbeault and dual Dolbeault cohomologies as depicted below
\begin{center}
\hspace{-.1cm}
{\vspace{-3cm}\epsfig{file=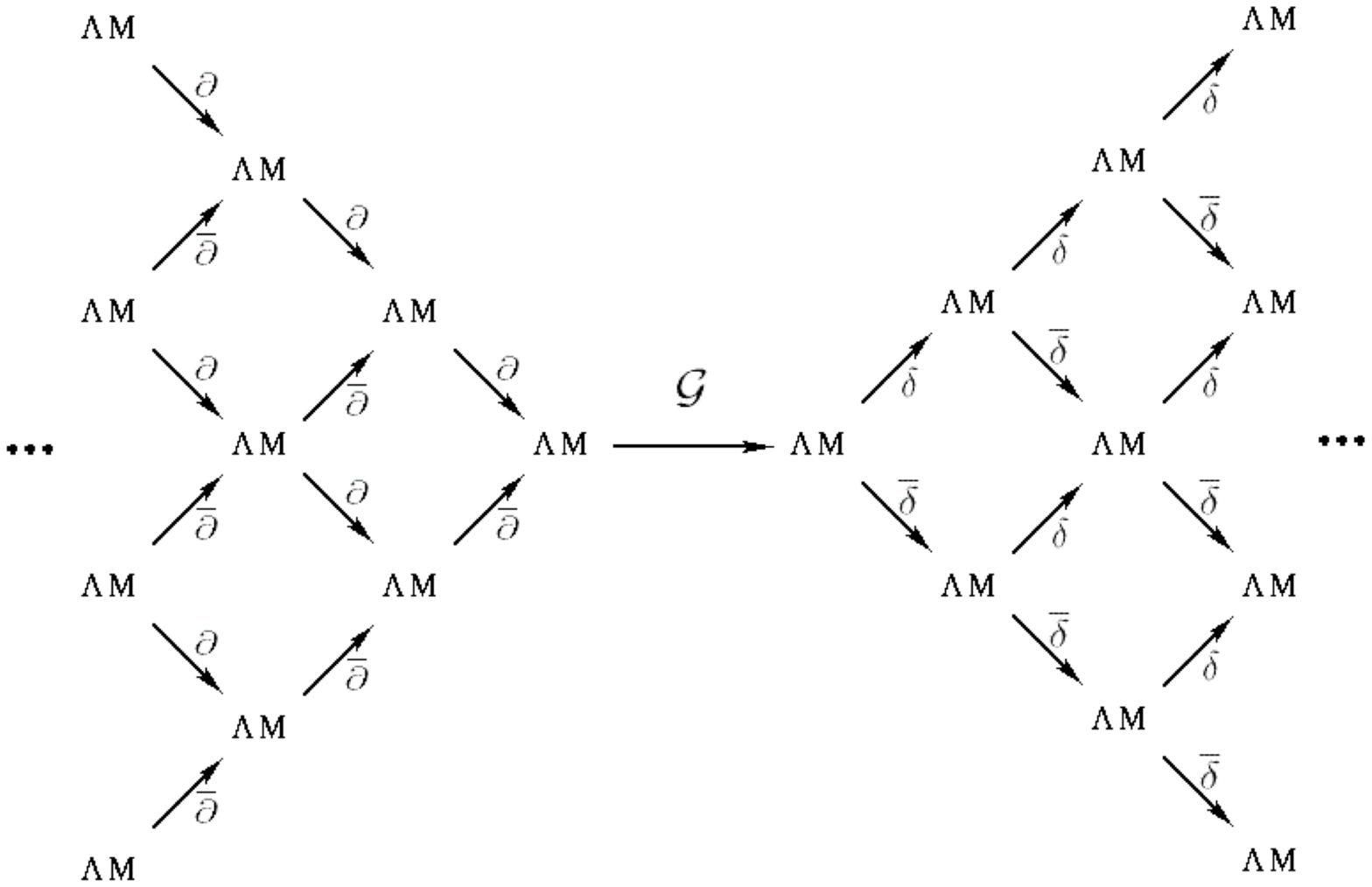,height=11cm,width=11cm}}
\end{center}
and determines a gauge invariant action principle~$S=\int_M (\varphi,{\cal G}\varphi)$
which describes a theory of~$(p,q)$-forms on a  K\"ahler manifold.
We stress that the equation of motion $\La {\bm \partial} \bar {\bm \partial} \psi_0=0$ is equivalent to~${\cal G}\psi_0=0$, as analyzed in the Note~\cite{Cherney:2009vg}. For the case where~$\psi_0$ is a 1-form in four dimensions, this is the linearized version of the holomorphic topological Yang-Mills theory studied in~\cite{Galperin:1990nm,Park}. In other dimensions, and for higher forms we refer to this new theory as~$(p,q)$-form
K\"ahler electromagnetism. 
Let us briefly highlight some of its properties:

Firstly, there is a strong analogy between the~${\cal N}=4$ supersymmetric K\"ahler model equipped with an~$\frak{sl}(2)$ Lefschetz symmetry and the symmetric tensor model with an~$\frak{sp}(2)$ $R$-symmetry. To see this, we use the algebra~(\ref{Kalg1},\ref{Kalg2}) 
to rewrite the equations of motion as
\be\nn
\Big[\bm\Delta - 2 {\bm\partial} {\bm \delta} -2 \bar{\bm \partial}\bar{\bm \delta}+2{\bm \partial}\bar{\bm \partial}\, \bm \Lambda\Big] \psi_0 = 0\, ,\label{maxeom}
\ee
which should be contrasted with~\eqn{singlecompdet}. Clearly the operators
~$(\bm\partial,\bar{\bm\partial})$, $(\bm \delta,\bar{\bm \delta})$,~$\L$, and~$ \La$ play the roles 
of~${\bf div}$,~$\GRAD,~\G$ and~$\TR$, respectively.
Further, the invertible operator
$:I_{1}(2\sqrt{\bf L\Lambda})/\sqrt{\bf L\Lambda}:$ 
plays the same role as $1-\frac14 \G\TR$; they promote equations of motion on a single field to equivalent equations of motion involving a self adjoint long operator with manifest gauge symmetries and Bianchi identities. A generalization of this construction to models with generalized supersymmetry algebras~\eqn{alg} and $R$-symmetry algebra $osp(p,q|2r)$ will be explored in~\cite{CLW}. 

\section{Conclusions}\label{conc}
BRST detour quantization is a method to construct gauge invariant  field theories from 
quantum mechanical constraint algebras. Its virtues include: 
(i)~Rather than starting from a known gauge theory, its input data consists only of a quantum constraint algebra.
(ii) The output data is a gauge invariant theory derivable from an action principle.
(ii) Given a quantum constraint algebra, the spectrum of the  {\it na\"ive} Dirac quantization need not coincide with the {\it desired} physical spectrum. The latter is encapsulated by the BRST detour quantization method. 
(iii) BRST cohomologies at all ghost numbers have simple physical interpretations as either solutions to gauge invariant equations of motion, gauge and gauge for gauge symmetries or Bianchi identities and ``Bianchi for Bianchi'' identities.

In this Article, we developed the BRST detour quantization method for superalgebras of the form~$[Q^*_i,Q^j\}=\delta^j_i{\cal D}$ with a single central generator ${\cal D}$. This lead to models with towers of physical fields obtained by expanding the zero ghost number BRST Hilbert space in powers of ghost bilinears~$y^{ij}$ (which themselves have zero ghost numbers).
These bilinears and their adjoints form a representation of the $R$-symmetry algebra on the ghost manifold. There is also a representation in terms of bilinears in the supercharges $(Q^*_i,Q^i)$.
The long operator, which determines the gauge invariant equations of motion, is obtained 
by building an $R$-symmetry invariant in the product of these representations. This result is borne out by our K\"ahler example and details will be provided in an upcoming publication~\cite{CLW}.
Models with a single field can be obtained by a ``ghostbusting'' procedure, which amounts to gauging the $R$-symmetries associated 
with the ghost bilinears $y^{ij}$. Again, the general result is similar to that found for our K\"ahler model and details will be given in~\cite{CLW}.

There is a deep connection between our approach and the pure spinors of Berkovits~\cite{Berkovits:2000fe} applicable to more general spacetime superalgebras.
Indeed, the ghost bilinears $y^{ij}$ can be viewed from that perspective as pure spinors. Although they are not spinors in spacetime,
their role producing a nilpotent operator from the supercharges is the same. Also, from a representation theoretic viewpoint, 
their origin as solutions to equations bilinear in phase space coordinates ({\it i.e.}, as elements of the kernel of the operator~$M$)
is a natural generalization of pure spinors (see~\cite{Pioline:2003bk}). This implies that our method can be applied to a range of models beyond the superalgebras considered here.

Going beyond rank~1 constraint algebras, we would like to study constraint algebras that close at only quadratic or higher order in the 
generators There are several motivations for this:
In~\cite{Bastianelli:2008nm} it is shown that  the first class constraint algebra of a spinning particle in a conformally flat background closes only quadratically. In constant curvature backgrounds, Dirac quantization yields a  Hilbert space spanned by generalized, conformal, 
spin~$s$ curvatures. Moreover a de Rham-like complex has  been constructed and the compensated equations of motion follow from tracing higher spin generalized curvatures. 
Another example of rank~2 algebras are those 
arising from the supersymmetric orthosymplectic quantum mechanical models of~\cite{Hallowell:2007qk}. In constant curvature backgrounds,
the superalgebras of these models have quadratic corrections built from the Casimirs of their $R$-symmetry generators.  
BRST charges for these types of models have been analyzed in~\cite{Buch}, so a detour quantization should again be possible.

Our final example was the original motivation for the work presented in this Note:
studies of black holes in four dimensional  ${\cal N}=2$ supergravities lead to 
spinning particles with~${\cal N}=4$ supersymmetries on the worldline, propagating on a quaternionic 
K\"ahler background~\cite{Gunaydin:2007bg}. To study their BRST quantization involves first class constraint algebras with structure functions
related to the underlying quaternionic K\"ahler structures. We have found a geometric construction of the BRST charges of these models
and will present their quantization in future work.

\section{Acknowledgements}

This work was partially supported by an NSF VIGRE award $\#$ DMS-0636297 and could not have been completed without extensive discussions with our colleagues. We are particularly grateful to
Boris Pioline, Martin Ro$\check{\rm c}$ek, Andy Neitzke, Maxim Grigoriev, Dmitry Fuchs, Albert Schwarz, Fiorenzo Bastianelli and
Olindo Corradini.

\end{document}